\documentclass{article}
\usepackage[top=2.5cm,bottom=2.5cm,left=2.5cm,right=2.5cm]{geometry}
\usepackage{amsmath}
\usepackage{amsfonts}
\usepackage{amssymb}
\usepackage{graphicx}
\usepackage{soul}
\usepackage{float}
\usepackage{color}
\usepackage[numbers]{natbib}
\usepackage{authblk}
\usepackage[colorlinks=true]{hyperref}

\newcommand{\bt}{\beta}

\newcommand{\G}{\Gamma}

\newcommand{\la}{\lambda}

\renewcommand{\part}{\partial}

\newcommand{\br}{{\mathbb R}}

\newcommand{\aver}[1]{\left\langle #1 \right\rangle}

\newtheorem{theorem}{Theorem}[section]
\newtheorem{example}{Example}[section]
\newtheorem{exercise}{Exercise}[section]

\newtheorem{lemma}{Lemma}[section]
\newtheorem{remark}{Remark}[section]

\newtheorem{proposition}{Proposition}[section]
\newtheorem{corollary}{Corollary}[section]
\newtheorem{definition}{Definition}[section]

\def\bt{\begin{theorem}}
\def\et{\end{theorem}}
\def\bc{\begin{corollary}}
\def\ec{\end{corollary}}
\def\bx{\begin{example}\small}
\def\ex{\end{example}}
\def\bxr{\begin{exercise}\small}
\def\exr{\end{exercise}}
\def\bl{\begin{lemma}}
\def\el{\end{lemma}}
\def\bd{\begin{definition}}
\def\ed{\end{definition}}
\def\bp{\begin{proposition}}
\def\ep{\end{proposition}}

\def\br{\begin{remark}}
\def\er{\end{remark}}

\def\be{\begin{equation}}
\def\ee{\end{equation}}
\def\bea{\begin{eqnarray}}
\def\eea{\end{eqnarray}}
\def\beas{\begin{eqnarray*}}
\def\eeas{\end{eqnarray*}}

\def\1{{\bf 1}}

\numberwithin{equation}{section}

\title{Spontaneous modulational instability of elliptic periodic waves: \\ the soliton condensate model}

\author[1,2,3]{D.\,S.~Agafontsev}
\author[1]{T.~Congy}
\author[1]{G.\,A.~El}
\author[4]{S.~Randoux}
\author[1]{G.~Roberti}
\author[4]{P.~Suret}

\affil[1]{Department of Mathematics, Physics and Electrical Engineering, Northumbria University, Newcastle upon Tyne, NE1 8ST, United Kingdom}
\affil[2]{Shirshov Institute of Oceanology of RAS, 117997, Moscow, Russia}
\affil[3]{Skolkovo Institute of Science and Technology, 121205, Moscow, Russia}
\affil[4]{Univ. Lille, CNRS, UMR 8523 - PhLAM -  Physique des Lasers Atomes et Mol\'ecules, F-59000 Lille, France}

\date{\today}

\begin{document}

\maketitle

\begin{abstract}
We use the spectral theory of soliton gas for the one-dimensional focusing nonlinear Schr\"odinger equation (fNLSE) to describe the statistically stationary and spatially homogeneous integrable turbulence emerging at large times from the evolution of the spontaneous (noise-induced) modulational instability of the elliptic ``dn'' fNLSE solutions. 
We show that a special, critically dense, soliton gas, namely the  genus one bound-state soliton condensate, represents an accurate model of the asymptotic state of the ``elliptic'' integrable turbulence. This is done by first analytically evaluating the relevant spectral density of states which is then used for   implementing  the soliton condensate numerically via a random $N$-soliton ensemble with $N$ large. A comparison of the statistical parameters, such as the Fourier spectrum, the probability density function of the wave intensity and the autocorrelation function of the intensity, of the soliton condensate with the results of direct numerical fNLSE simulations with $\mathrm{dn}$ initial data augmented by a small statistically uniform random perturbation (a noise) shows a remarkable agreement. Additionally, we analytically compute the kurtosis of the elliptic integrable turbulence, which enables one to estimate the deviation from Gaussianity. The analytical predictions of the kurtosis values, including the frequency of its temporal oscillations at the intermediate stage of the modulational instability  development, are also shown to be in excellent agreement with numerical simulations for the entire range of the elliptic parameter $m$ of the initial $\mathrm{dn}$ potential. 
\end{abstract}


\section{Introduction}
\label{Sec:Intro}

Modulational instability (MI) is a fundamental physical phenomenon that has been attracting a major attention from various physics and mathematics communities  over the last six decades~\cite{Zakharov:MI:09}. Pioneered by  Whitham, Lighthill, Benjamin and Feir,  Ostrovsky, and Zakharov in 1960-s the theory of modulational instability has developed  into a broad area of research with numerous applications in water waves \cite{Osborne}, nonlinear optics \cite{Agrawal:2013},  and condensed matter physics \cite{strecker_formation_2002}. Typically, MI is manifested  as  a temporal growth of the amplitude  of a small perturbation (modulation) of a weakly nonlinear periodic  wave.  
In many scenarios the  dispersion of the  wave's envelope plays the dominant role at the initial (linear) stage of the MI development. 
The initial exponential growth is saturated by nonlinear effects when the modulation amplitude becomes sufficiently large, leading at longer times to the emergence  of coherent structures such as solitons and breathers. The eventual fate of the modulationally unstable periodic wave strongly depends on the type of dynamics (integrable vs non-integrable) and the shape of the initial perturbation  (localised vs periodic vs random, cf. \cite{akhmediev1986modulation,
Zakharov:13, agafontsev2015integrable,  agafontsev2021rogue, Grinevich:18, Bonnefoy2020modulational, Biondini:16:prlmi}). Some scenarios of the MI development  involve the generation of rogue waves---localised coherent structures of unusually large amplitude that emerge  unpredictably within otherwise moderate-amplitude wave landscape \cite{kharif2008Rogue}. 

 It has been well recognised that the  one-dimensional cubic focusing NLS equation (fNLSE) represents a paradigmatic model for the description of MI of weakly nonlinear narrow-band short waves (also known as Stokes waves). In the standard normalized form the fNLSE is represented as
\begin{equation}\label{NLSE}
	i\psi_t + \psi_{xx} + 2|\psi|^2 \psi=0, \ \ \psi \in \mathbb{C},
\end{equation}
where $\psi(x,t)$ is the wave envelope, $t$ is the time-like variable and $x$ -- the space-like variable (depending on the application, the $t$-variable in \eqref{NLSE} can have the meaning of a physical spatial variable, e.g. the propagation length in optical fibres, while the $x$-variable corresponds to the physical time). There have been numerous analytical, numerical and experimental investigations of various aspects of MI in physical systems modeled by the fNLSE (see \cite{dafermos_geometry_1986,
tracy_nonlinear_1988,kamchatnov1997new, akhmediev1986modulation,
Zakharov:13, agafontsev2015integrable,  agafontsev2021rogue, Grinevich:18, Bonnefoy2020modulational, Biondini:16:prlmi} and references therein). The central role in these studies is played by the integrability of the fNLSE in the framework of the inverse scattering transform (IST) developed by Zakharov and Shabat (ZS) \cite{Zakharov:1972Exact}. The IST associates the fNLSE evolution with a pair of two linear problems (the so-called Lax pair) one of which is the scattering/eigenvalue  problem for the non-self-adjoint Dirac (ZS) operator in which  fNLSE solution $\psi(x,t)$ plays the role of the potential. 

Of particular significance for the theory and applications of MI are two types of exact solutions to the fNLSE \eqref{NLSE}: the plane (continuous) waves described by the spatially uniform  time-periodic exponential solution
\be \label{pw}
    \psi = q\,e^{2iq^{2}t}, \ \ q>0,
\ee 
and  nonlinear spatially periodic waves described by the Jacobi elliptic functions --  the  ``cn'' family and the ``dn''family. In the literature both families of elliptic solutions are often called ``cnoidal waves'' and they are classified as genus one fNLSE solutions within the finite-gap theory (the plane waves \eqref{pw} are genus zero solutions)~\cite{Osborne, novikov1984theory, tracy_nonlinear_1988}.  
At the same time, the $\mathrm {dn}$ and $\mathrm {cn}$ solutions have 
qualitatively different configurations of their ZS spectrum  with the (finite-gap) spectrum of dn being located along the imaginary axis while the spectrum of cn solution lying along two Schwartz-symmetric arcs in $\mathbb C$~\cite{Osborne, tracy_nonlinear_1988}. Both families are modulationally unstable with respect to small perturbations~\cite{kuznetsov1999modulation,DECONINCK2017}.  In this paper we focus on the dn family only.

The notion of integrable turbulence, both as a physical phenomenon and as a theoretical framework in which to study random solutions to integrable wave equations, was introduced by V. Zakharov in 2009 \cite{Zakharov:09}. The relation between MI and integrable turbulence was first explored in \cite{agafontsev2015integrable} where the long-time development of a random noise perturbation of a plane wave fNLSE solution was studied numerically. It was shown that the  development of  spontaneous MI of a plane wave  results, in the long time regime,  in the emergence of strongly nonlinear statistically stationary integrable turbulence characterised by Gaussian single-point statistics and very peculiar behaviours of the Fourier power spectrum and the autocorrelation function~\cite{kraych2019statistical}. The numerical results of~\cite{agafontsev2015integrable} were extended in~\cite{agafontsev2016integrable} to the case of the noise-induced MI of {\it elliptic periodic} solutions of fNLSE. 
It was shown in~\cite{agafontsev2016integrable} that, similarly to the spontaneous MI of plane waves, the integrable turbulence developed in the long-time evolution of nonlinear elliptic $\mathrm {dn}$ solutions augmented by a small noise is characterized by stationary statistical distributions, whose concrete form depends on the elliptic parameter $m$ of the initial  $\mathrm {dn}$ wave. Specifically, the long-time asymptotic turbulent field was shown to exhibit Gaussianity when $m \to 0$, i.e. when the $\mathrm {dn}$ function approaches the plane wave, and to display strong deviations from Gaussian statistics as $m$ departs from $0$.

A {\it soliton gas} model of fNLSE integrable turbulence emerging from the development of spontaneous MI of plane waves was proposed in \cite{gelash2019bound}. It was shown numerically that the statistical parameters of long-time integrable turbulence (the probability density function, the power spectrum, the autocorrelation of intensity) agree with high accuracy  with those of a {\it soliton condensate} --- a
criticaly dense  random $N$-soliton ensemble with $N \gg 1$,  characterized by the so-called Weyl distribution over the IST spectrum and with the soliton phases (the phases of the norming constants in the $N$-soliton) being  independent random values, each distributed uniformly  on $[0, 2 \pi)$. The theoretical notion of soliton condensates as the special class of  soliton gases whose collective (emergent) dynamics are entirely dominated by soliton interactions with no individual soliton states discernible, was introduced in \cite{el_spectral_2020}. Various properties of soliton condensates were studied  in \cite{congy_dispersive_2023} for the Korteweg-de Vies (KdV) equation and in \cite{tovbis_periodic_2022} for the fNLSE. 

The kinetic theory of soliton gases as out of equilibrium infinite-soliton ensembles characterized by random amplitudes and positions/phases was initiated in 1971 by Zakharov \cite{zakharov1971kinetic} in the context of a low density (rarefied) gas of  KdV solitons. This theory has been later generalized to the case of dense KdV gases  \cite{el_thermodynamic_2003} and further, in \cite{el_kinetic_2005, el_spectral_2020}, to fNLSE soliton and breather gases. The key quantities describing soliton gas in terms of the IST spectrum $\lambda \in \mathbb{C}$ are the density of states (DOS) $f(\lambda)$ and the spectral flux density $g(\lambda)$, the respective analogues of the wavenumber and the frequency of a nonlinear periodic wave. These quantities satisfy the nonlinear dispersion relations (NDRs) first derived for the fNLSE soliton gas in \cite{el_spectral_2020}.  The NDRs are a pair of integral equations (see~\eqref{NDRs}) parametrized by the so-called spectral scaling function $\sigma (\lambda)$, which can be thought of as a certain analogue of temperature (see \cite{bonnemain_generalized_2022} for the statistical mechanics interpretation of the spectral theory of a soliton gas for the KdV equation).  The soliton condensate limit is spectrally realized in the NDRs by letting $\sigma \to 0$. As a result,  a soliton condensate  is uniquely  defined by its spectral support, i.e. the locus $\Lambda \subset \mathbb{C}$ of the soliton IST eigenvalues $\lambda$. Note that generally, the spectral support $\Lambda$ of a condensate can be composed of a finite number of disjoint 1D sets (line intervals, or arcs), $\Lambda= \gamma_1 \cup \gamma_2 \cup \dots \cup \gamma_{n+1}$, where $n$ is called {\it the genus} of the condensate. 
Importantly, the NDRs for soliton condensates can be solved  in an explicit form for certain configurations of the spectral support $\Lambda$, e.g. for the bound state fNLSE condensates with $\Lambda \in i \mathbb{R}$ or for the condensates with a circular spectral support $\Lambda = \{\lambda: |\lambda|=r>0\}$ \cite{el_spectral_2020}. The solution $f(\lambda), g(\lambda)$  of the NDRs can then be used for the evaluation of various physical observables in the soliton condensate.

Soliton condensates for the KdV equation  were shown in \cite{congy_dispersive_2023} to (almost surely) coincide with finite-gap potentials \cite{novikov1984theory, Osborne}. Such condensates can be viewed as ``regular'' or ``deterministic'' soliton gases, which were studied in \cite{girotti_rigorous_2021, girotti_soliton_2023} via asymptotic analysis of the so-called primitive potentials \cite{dyachenko2016primitive}.  From this perspective  the KdV soliton condensates can be viewed as the ``phase-locked'', macroscopically coherent dispersive-hydrodynamic states generalizing the notions of rarefaction and dispersive shock waves  \cite{el_dispersive_2016}. In contrast, the fNLSE soliton condensates are inherently random wave fields due to  the  random distribution of the soliton phases (which are independent of the soliton positions in space).  We note here that in the literature on fNLSE the term ``condensate'' is often applied  to the plane wave solution \eqref{pw}.  Within the  soliton gas framework  this solution represents a {\it particular  realization} of a genus zero bound state soliton condensate corresponding to a specially chosen set of soliton phases. Such solitonic approximation of the plane wave  was numerically realized in \cite{gelash2021solitonic}.
The ``full'' genus zero condensate equipped with a random phase distribution  was shown in \cite{gelash2019bound} to accurately model  the integrable turbulence resulting from the development of the spontaneous MI of a plane wave. 

Numerical implementation of a soliton gas is achieved via building $N$-soliton ensembles with $N\gg1$ and appropriately configured distributions for the soliton spectra and norming constants. An  effective algorithm  of the numerical synthesis of fNLSE soliton gas based on the Darboux transform was developed in \cite{gelash2018strongly} (see also~\cite{gelash2019bound}). 

In this paper we extend the results of \cite{gelash2019bound} on the solitonic model of MI of plane waves and  show that  genus 1 soliton condensates accurately describe  the integrable turbulence  of the long-time development of the spontaneous  MI of  the $\mathrm{dn}$ family of nonlinear periodic fNLSE solutions (see Eq. \eqref{dn-branch}). We compare the statistical parameters of the soliton condensate model with the results of direct numerical simulations of fNLSE and observe excellent agreement.
In particular, we evaluate analytically the fourth normalized moment of the wave field amplitude related to the kurtosis of the probability density function (PDF) in the soliton condensate and find that our analytical prediction is in excellent agreement with the corresponding value extracted from the numerical simulations of the spontaneous MI development. The obtained dependence of the kurtosis $\kappa_4$ of the soliton condensate on the elliptic parameter $m$ of the input $\mathrm{dn}$ solution shows that $\kappa_4>2$ for $m>0$ indicating a heavy-tailed PDF and implying the presence of rogue waves in the developed ``elliptic'' integrable turbulence in agreement with earlier numerical results of \cite{agafontsev2016integrable}. The  limiting value $\kappa=2$ corresponds to the Gaussian statistics and is realized for $m=0$ in the  setting of this paper. This is also consistent with the previous results \cite{agafontsev2015integrable, gelash2019bound} on the spontaneous MI of plane waves since the case $m=0$ corresponds to the degeneration of the $\mathrm{dn}$ fNLSE solution \eqref{dn-branch} to the plane wave solution \eqref{pw}. In this connection we mention a  related recent study \cite{biondini_breather_2024} where the evolution of the elliptic data $\psi(x,0)=\mathrm{dn}(x;m)$ augmented by a small noise perturbation was studied in the small-dispersion (semi-classical) fNLSE framework. In that case the $\mathrm{dn}$ wave  \eqref{dn-branch} is not an exact solution of the (small-dispersion) fNLSE, and the resulting integrable turbulence is not modeled by a ``pure'' soliton condensate as in the present paper but was shown to be approximated by a  breather gas with a  complex spectral structure.

The structure of the paper is as follows. In Section~\ref{Sec:Sec2} we provide a brief summary of the necessary results on MI, integrable turbulence and soliton gases with a particular emphasis on the spectral properties of soliton condensates. Section~\ref{Sec:Sec3} is central and is devoted to the comparison of the statistical  parameters of the genus one soliton  condensate with the counterpart parameters obtained from the direct fNLSE numerical simulation of the long-time development of the MI of elliptic $\mathrm{dn}$ wave  augmented by random noise of small amplitude. In Section~\ref{Sec:Conclusions} we present conclusions and outloook of our study. The Appendix contains some technical details on the multisoliton solutions of the fNLSE and the description of the numerical methods used.


\section{MI, Integrable turbulence and soliton condensates}
\label{Sec:Sec2}


\subsection{MI of cnoidal waves and integrable turbulence}
\label{Sec:Sec2-1}

Cnoidal waves are exact periodic solutions of the fNLSE~(\ref{NLSE}), which are expressed in terms of the elliptic functions and depend essentially on two parameters, e.g. the real and imaginary half-periods, $\omega_{0}$ and $\omega_{1}$ respectively~\cite{kuznetsov1999modulation, a_m_kamchatnov_nonlinear_2000}. 
While there are dn- and cn-branches of such solutions, in the present paper we study only the dn-branch, which is written as 
\begin{eqnarray}
	\psi_{\mathrm{dn}}(x,t) = e^{i\Omega t}\,\nu\,\,\mathrm{dn}(\nu x; m),
	\label{dn-branch}
\end{eqnarray}
where $\mathrm{dn}(x; m)$ is the so-called delta amplitude Jacobi elliptic function, while  $\Omega$, $\nu$ and $m$ are expressed in terms of the two chosen parameters of the $\mathrm{dn}$, e.g. $\omega_{0}$ and $\omega_{1}$ (a convenient spectral parametrization of the cnoidal wave solution will be presented below). The elliptic parameter $m \in [0,1]$ controls the waveform of~\eqref{dn-branch}: from a vanishing amplitude harmonic wave on a nonzero plane wave background~\eqref{pw} as $m \to 0$ ($\omega_1/\omega_0 \to +\infty$ for fixed $\omega_{0}$) to a  solitary wave on a zero background when $m \to 1$ ($\omega_1/\omega_0 \to 0$). 

Within the IST  framework the solution \eqref{dn-branch} is characterized by the finite-gap spectrum of the Lax (Zakharov-Shabat) operator in the linear scattering problem associated with the fNLSE equation \cite{Zakharov:1972Exact}. Namely, the Zakharov-Shabat spectrum of the potential \eqref{dn-branch} consists of two Schwarz-symmetric disjoint intervals (bands) $[-i\eta_2, - i\eta_1] \cup [i\eta_1, i \eta_2]$ on the imaginary axis. It is thus a one-gap potential, also called a genus 1 potential in the finite-gap theory \cite{novikov_theory_1984, Osborne}. The parameters of the $\mathrm{dn}$ solution \eqref{dn-branch} are expressed in terms of the spectral band edges $i\eta_1, i\eta_2$ as follows (see e.g. \cite{a_m_kamchatnov_nonlinear_2000}):
\be \label{nu_m}
\nu = \eta_1+ \eta_2, \quad m=\frac{4\eta_1 \eta_2}{(\eta_1 + \eta_2)^2}\, , \ \ 
\Omega=2(\eta_1^2 +\eta_2^2)\, .
\ee
Solution~(\ref{dn-branch}) is periodic in space with period 
\be
    2\omega_{0}=2\frac{ K(m)}{\eta_1 + \eta_2},
\ee
where $K(m)$ is the complete elliptic integral of the first kind.
At $t=0$, expression~\eqref{dn-branch} is real-valued and positive; see the example with half-periods $\omega_{0}=\pi$ and $\omega_{1}=1.6$ ($m\approx 0.48$, $\eta_1\approx 0.41$, $\eta_2\approx 0.59$) shown with the black line in Fig.~\ref{fig:figCW}. 
Note that, without loss of generality, in our numerical simulations we will consider cnoidal waves with the fixed real half-period $\omega_{0}=\pi$: indeed, the fNLSE can be rescaled in space, time and amplitude, making it possible to fix the real half-period, as well as the dispersion and nonlinearity coefficients as in~(\ref{NLSE}). 
We add that the fixed half-period $\omega_0=\pi$ was used in the simulations of~\cite{agafontsev2016integrable} so we shall also use this value here to make comparisons with previous  results. 

Solution \eqref{dn-branch} is modulationally unstable for $m>0$; the maximum growth rate of the MI was found in~\cite{kuznetsov1999modulation}, and we express it here in terms of the spectral parameters,
\begin{eqnarray}
	\gamma_{\max} = 2\nu^{2}\,\sqrt{1- m} = 2[\eta_2^2 - \eta_1^2].
	\label{gamma_max}
\end{eqnarray}
When $m\to 1$ (equivalently $\eta_1 \to \eta_2$), we have $\gamma_{\max} \to 0$, which corresponds to the modulational stability of the fundamental fNLSE solitons. 
When $m\to 0$ ($\eta_1 \to 0$), we have $\gamma_{\max} \to 2 \eta_2^2$, which corresponds to the growth rate of the most unstable harmonic perturbation of the plane wave solution~\eqref{pw} with $q=\eta_2$.

The statistical properties of the spontaneous (noise-induced) MI developing from the dn-branch of cnoidal waves have been studied numerically in~\cite{agafontsev2016integrable}. 
It has been observed that, after a transient evolution of the statistical functions in the form of damped oscillations, the integrable turbulence emerging from the MI approaches asymptotically its statistically stationary state with time. 
Further, for the statistical functions such as the normalized fourth-order moment of wave amplitude $|\psi|$ (we shall call it kurtosis with a slight abuse of the terminology) $\kappa_{4} = \langle|\psi|^{4}\rangle/\langle|\psi|^{2}\rangle^{2}$, the amplitude of the transient oscillations decays with time by a power law, the phase contains the nonlinear phase shift decaying as $t^{-1/2}$, and the frequency of the oscillations $s$ turns out to be very close to the double maximum growth rate of the MI, $s\approx 2\gamma_{\max}$, see the formula~\eqref{oscillations-ansatz} below.

The asymptotic stationary values of the statistical parameters, e.g., the kinetic and potential energies, the Fourier spectrum and the probability density function (PDF) of wave intensity $|\psi|^{2}$, have been computed. 
It turned out that, in the long time, the potential energy is twice as large as the kinetic one, the Fourier spectrum contains peaks at certain wavenumbers near which it diverges by power law, and the PDF is close to the exponential distribution for cnoidal waves with $m\to 0$ (the plane wave limit), and is significantly non-exponential for cnoidal waves with $m$ sufficiently close to $1$. 


\subsection{Spectral theory of fNLSE soliton gas and soliton condensates}
\label{Sec:Sec2-2}

Motivated by the results of Ref.~\cite{gelash2019bound}, where the long-time development of the spontaneous MI of plane waves  was acutely modeled by a dense soliton gas (specifically, the genus 0 bound state soliton condensate), in this paper we introduce the solitonic model of the spontaneous MI of cnoidal waves \eqref{dn-branch}. Specifically, we consider the genus 1 soliton condensate with the same spectral support as the dn solution spectrum but with randomly distributed soliton phases. This generalization is non-trivial because the determination of the density of states of the ``elliptic'' soliton gas corresponding to the $\mathrm{dn}$ solution \eqref{dn-branch} involves solution of  the integral nonlinear dispersion relation for soliton condensate while for the case of the MI of plane wave the associated soliton gas DOS can be simply extracted from the semi-classical Bohr-Sommerfeld distribution for the box potential as it was done in \cite{gelash2019bound}. To justify our soliton condensate model  we will  compare in Section~\ref{Sec:Sec3} the statistical characteristics of the developed noise-induced MI of the  dn solution with the same spectral characteristics of the associated genus 1 soliton condensate built numerically as an appropriately configured dense $N$-soliton ensemble with $N\gg 1$.  In the limit $m\to 0$ our results are expected to reduce to the previous results of \cite{gelash2019bound} for MI of plane waves.


\subsubsection{Soliton gas: density of states and nonlinear dispersion relations}

The concept of soliton gas (SG) as an infinite-soliton ensemble characterized by random amplitudes and positions/phases was introduced in 1971 by Zakharov \cite{zakharov1971kinetic} in the context of a low density (rarefied) gas of  KdV solitons. This theory has been later  generalized to the case of dense KdV SGs  \cite{el_thermodynamic_2003} and further, in \cite{el_kinetic_2005, el_spectral_2020}, to fNLSE soliton and breather gases. In these works SG was defined in terms of infinite-genus, thermodynamic limits of finite-gap potentials generalizing the multisoliton solutions. In the recent work \cite{jenkins_approximation_2024} the  multisoliton approximation of fNLSE SGs  was rigorously  established via  the Riemann-Hilbert approach, thus reconciling  the two existing approaches to SGs.
We refer the reader to the  reviews \cite{el_soliton_2021, suret_soliton_2024} for a detailed exposition of the SG theory and applications.

The key quantities describing fNLSE SG in terms of the IST spectrum $\lambda \in \mathbb{C}$ are the density of states (DOS) $f(\lambda)$ and the spectral flux density $g(\lambda)$ \cite{suret_soliton_2024}. The DOS $f(\lambda)$ of a spatially homogeneous, equlibrium SG is phenomenologically defined as a number of soliton states contained in a unit element of the spectral phase space, i.e. the joint density distribution of the soliton spectral parameters on some compact $\Lambda \subset \mathbb{C}$ and their positions on $\mathbb{R}$. 
The complementary SG characteristic, the spectral flux density $g(\lambda)$ is defined for a non-bound state SG (i.e. a gas in which solitons have different nonzero velocities) as the number of solitons per unit element of the spectral parameter space, crossing any given point $x=x_0$ per unit interval of time. 
For the bound state fNLSE SG $g(\lambda) \equiv 0$.

The rigorous  definitions of the DOS and the spectral flux density are achieved in the framework of the thermodynamic limit of finite-gap potentials \cite{el_spectral_2020} where  $f(\lambda)$ and $g(\lambda)$ are shown to satisfy certain {\it Nonlinear Dispersion Relations} (NDRs).  We introduce the Schwarz-symmetric spectral support $\Lambda \subset \mathbb{C}$ of the SG DOS, and consider only the upper half plane so that $\Lambda^+=\Lambda \cap \mathbb{C}^+$. Then the NDRs for the fNLSE SG have the form \cite{el_spectral_2020}
\begin{equation} \label{NDRs}    \begin{split}
& \int _{\Lambda^+}\ln \left|\frac{z-\bar\la}{z-\la}\right|
f(z)d \alpha (z)+\sigma(\la)f(\la) =  \text{Im}\, \la,  \\
& \int _{\Lambda^+}\ln \left|\frac{z-\bar\la}{z-\la}\right| g(z)d \alpha (z) +  \sigma(\la) g(\la) = -4\, \text{Im}\, \la \, \text{Re}\,\la, 
 \end{split}
\end{equation}
where $d\alpha (z)$ is a reference measure, e.g. the arclength $d \alpha(z) = |dz|$  if $\Lambda^+$ is a 1D curve in $\mathbb{C}^+$.
The function $\sigma(\lambda) >0$ is the so-called spectral scaling function, which characterizes the  spectral Riemann curve of the finite gap potentials in the thermodynamic limit \cite{el_spectral_2020}.  For weakly non-homogeneous SG $f(\lambda) \to f(\lambda;x,t)$, $g(\lambda) \to g(\lambda;x,t)$, and the NDRs are complemented by the kinetic equation $f_t + g_x=0$ \cite{el_soliton_2021, suret_soliton_2024}.

For the bound state non-propagating soliton gas $\Lambda^+ \subset i\mathbb{R}^+$ so that we have $\lambda = i \eta$, $z=i\mu$, $\eta>0$, $\mu>0$, and the second NDR is identically satisfied by $g(\lambda)\equiv 0$.


\subsubsection{Bound state soliton condensates}
\label{sec:bound}

Soliton condensates can be viewed as  critically dense
SGs  constrained by a given spectral support $\Lambda^+$ \cite{el_spectral_2020}.
Spectrally they are realized by letting $\sigma \to 0$ in the NDRs \eqref{NDRs}. 
Then in the condensate limit for the bound state gas the first NDR in \eqref{NDRs} assumes the form
\begin{equation} \label{NDR_cond}
  \int_{\Gamma^+} \ln \left|\frac{\mu+\eta}{\mu-\eta}\right|
  f(\mu)d\mu = \eta,
\end{equation}
where $\Gamma^+=[\eta_1, \eta_2]\cup [\eta_3, \eta_4]\cup \dots \cup [\eta_{2g-1}, \eta_{2g}]$. Here $g\in \mathbb{N}$ is the genus of the condensate. For $g=0$ we define $\Gamma^+=[0, \eta_1]$.

We now observe that the integral equation \eqref{NDR_cond} coincides with the NDR for the genus $g$ soliton condensate for the KdV equation (see Ref.~\cite{congy_dispersive_2023} where the spectral theory for KdV soliton condensates has been constructed). For genus zero its  solution  is given by
\be \label{weyl}
f(\eta)=f^{(0)}(\eta; \eta_1) = \frac{\eta}{\pi\sqrt{\eta_1^2-\eta^2}},\ee
which is the so-called Weyl distribution used in \cite{gelash2019bound} for the modelling of the spontaneous MI of plane waves. 
\begin{figure}[t]\centering
	\includegraphics[width=0.4\linewidth]{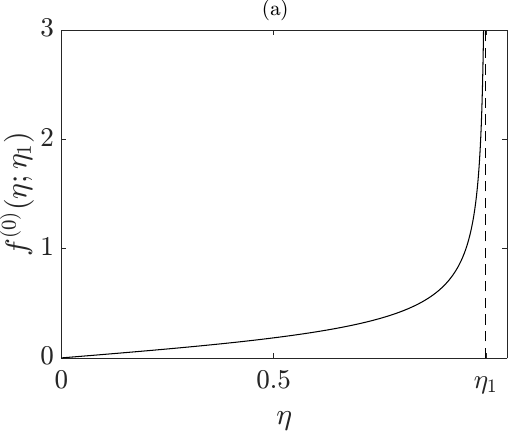}\hspace{1.5cm}\includegraphics[width=0.4\linewidth]{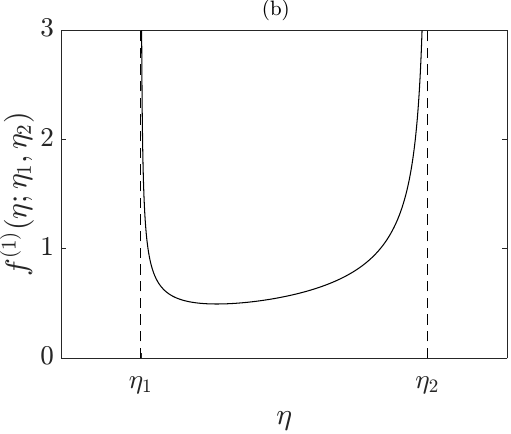}
	\caption{\small (a) DOS of genus $0$ condensate \eqref{weyl} with $\eta_1=1$. (b) DOS of genus $1$ condensate~\eqref{DOS_cond} with $\eta_1 = 0.41$ and $\eta_2 = 0.59$. }
	\label{fig:genus0}
\end{figure}

For the genus $1$ bound state  condensate with $\Gamma^+=[\eta_1, \eta_2]$ the  solution of \eqref{NDR_cond} assumes the form \cite{congy_dispersive_2023} 
\be \label{DOS_cond}
f(\eta) = f^{(1)}(\eta;\eta_1,\eta_2) = \frac{\eta^2-w^2}{\pi \sqrt{(\eta^2-\eta_1^2)(\eta_2^2-\eta^2)}}, 
    \ee
    where
    \be \label{w}
  w^2 = \eta_2^2\left(1-\frac{E(\tilde m)}{K(\tilde m)} \right),\quad \tilde m=\frac{\eta_1^2}{\eta_2^2}. 
\ee
Here  $E(\tilde m)$ is the complete elliptic integral  of the 2nd kind.
Note that the  parameter $\tilde m \in [0,1] $ in \eqref{w} does not coincide with the elliptic parameter $m$ in the dn solution \eqref{dn-branch} given by \eqref{nu_m}. Typical plots of the soliton condensate DOS's \eqref{weyl} and \eqref{DOS_cond} are  displayed in Fig.~\ref{fig:genus0}.

For ensemble averages in a general bound-state SG the following general formulae  were obtained in \cite{congy_statistics_2024}:
\begin{eqnarray}
    \aver{|\psi|^2} = \int_{\G^+} 4\eta f(\eta) d\eta, \quad
    \aver{|\psi|^4} = \int_{\G^+} \frac{32}{3}\eta^3 f(\eta) d\eta.
    \label{moments24a}
\end{eqnarray}
For the case of genus 1 condensate with $\Gamma^+=[\eta_1, \eta_2]$ and the DOS \eqref{DOS_cond} we obtain from \eqref{moments24a} (see Appendix~\ref{Sec:Integrals} for some useful integrals)
\begin{eqnarray}
 \aver{|\psi|^2} = \eta_1^2+\eta_2^2 - 2w^2, \quad
\aver{|\psi|^4} = 2(\eta_1^4+\frac23 \eta_2^2 \eta_1^2+ \eta_2^4 - \frac{4}{3} (\eta_1^2+\eta_2^2)w^2),\label{moments24}
\end{eqnarray}
where $w(\eta_1, \eta_2)$ is given by \eqref{w}.



\section{Genus 1 soliton condensate vs MI of elliptic periodic waves}
\label{Sec:Sec3}

In this Section we shall use the DOS \eqref{DOS_cond} to numerically realize the  genus 1 soliton condensate via $N$-soliton ensemble with $N \gg 1$ and compare its statistical parameters with those obtained by direct numerical simulations of the fNLSE \eqref{NLSE} with the cnoidal wave initial condition \eqref{dn-branch} augmented by a small statistically uniform random perturbation (a noise).


\subsection{Solitonic model of an unperturbed cnoidal wave}
\label{Sec:Sec3-1}

We start with the construction of a $N$-soliton approximation of an unperturbed cnoidal wave \eqref{dn-branch}. Our construction is based on the conjecture that the elliptic solution \eqref{dn-branch} can be accurately approximated by a special realization of the genus 1 bound state soliton condensate with the DOS \eqref{DOS_cond} supported on $\Gamma^+=[\eta_1, \eta_2]$. This conjecture, motivated by the previous analysis of the KdV soliton condensates in \cite{congy_dispersive_2023} will be confirmed by the numerical implementation of the $N$-soliton bound state  solution with $N \gg 1$ and the  DOS  given by  \eqref{DOS_cond}. 

First, we introduce a set $\{ \lambda_{n}=i\zeta_{n}\,\, |\,\, n=1,...,N; \ \zeta_{l}<\zeta_{m}\,\,\mbox{for}\,\,l>m \}$ of soliton eigenvalues, sorted in \textit{descending order},  so that $\zeta_n$ are distributed on $[\eta_1, \eta_2]$ with the density $\phi(\zeta)$ defined by the normalized condensate DOS \eqref{DOS_cond}:
\begin{align}
 \label{phi}
\phi(\zeta) = \frac{f^{(1)}(\zeta; \eta_1, \eta_2)}{ \int_{\eta_1}^{\eta_2} f^{(1)}(\eta; \eta_1, \eta_2) d \eta}.
\end{align}
We then sample the values $\zeta_n$, $n=1, \dots, N$ from $[\eta_1, \eta_2]$ according to
\be \label{sample}
\int_{\eta_1}^{\zeta_n} \phi(\eta) d \eta = \frac{n}{N} .
\ee
To effectively implement the discretization \eqref{sample} we evaluate  the integrated density of states 
 \begin{align}
 \label{eq:U2}
 \begin{split}
     U(\zeta) &= \int^{\zeta}_{\eta_1} f^{(1)}(\eta;\eta_1,\eta_2)\,d\eta\\ &=\frac{\eta_2}{\pi}\bigg[E(k,1-\tilde{m})-\left(1-\frac{E(\tilde{m})}{K(\tilde{m})}\right)F(k,1-\tilde{m}) 
 -\frac{1}{\zeta\eta_2}\sqrt{(\eta_2^2-\zeta^2)(\zeta^2-\eta_1^2)} \bigg],
 \end{split}
\end{align}
where $F(k,\tilde m)$ is the incomplete elliptic integral of the first kind,  $k=\arcsin{\sqrt{\frac{\eta_2^2(\zeta^2-\eta_1^2)}{\zeta^2(\eta_2^2-\eta_1^2)}}}$, and $\tilde m$ is defined by \eqref{w}. Then \eqref{sample} assumes the form
\be \label{eigen}
\frac{U(\zeta_n)}{U(\eta_2)}=\frac{n}{N}, \ \ n=1, \dots, N.
\ee
To define the $N$-soliton solution ($N$-SS) the solitonic eigenvalues $\zeta_n$ must be complemented by the norming constants. 
Following~\cite{gelash2021solitonic}, we assign to these eigenvalues the following norming constants,
\begin{eqnarray}
	C_{n} = (-1)^{n}. \label{DM-norming-constants-corrected}
\end{eqnarray}
Note that here and below we use norming constants in the dressing method (DM) formalism rather than in the conventional IST formalism; see e.g.~\cite{aref2016control,gelash2020anomalous} for the difference. The $N$-SS written in terms of DM norming constants is given in Appendix \ref{sec:nsoliton}. 
In the DM formalism, the norming constants are connected with the soliton position parameters $x_{0n}$ and phase parameters $\theta_{0n}$ (or simply positions and phases) as 
\begin{eqnarray}
	C_{n} = -\exp\bigg[2i\lambda_{n}x_{0n} + i\theta_{0n}\bigg], \label{Ck_param}
\end{eqnarray}
so that~(\ref{DM-norming-constants-corrected}) corresponds to all solitons sitting at the origin with the alternating phases of $0$ and $\pi$,
\begin{equation}
	x_{0n}=0, \quad \theta_{0n} = \frac{\pi}{2}[1 - (-1)^{n-1}]. \label{DM-positions-phases-box}
\end{equation}
Note that $x_{0n}$ and $\theta_{0n}$ coincide with the physical (observable) spatial position and the phase of a soliton only for one-soliton solution. 
In presence of other solitons or dispersive waves, the observed position and phase of a given soliton (if they can be accurately defined at all) may differ considerably from these parameters.

\begin{figure}[t]\centering
	\includegraphics[width=0.6\linewidth]{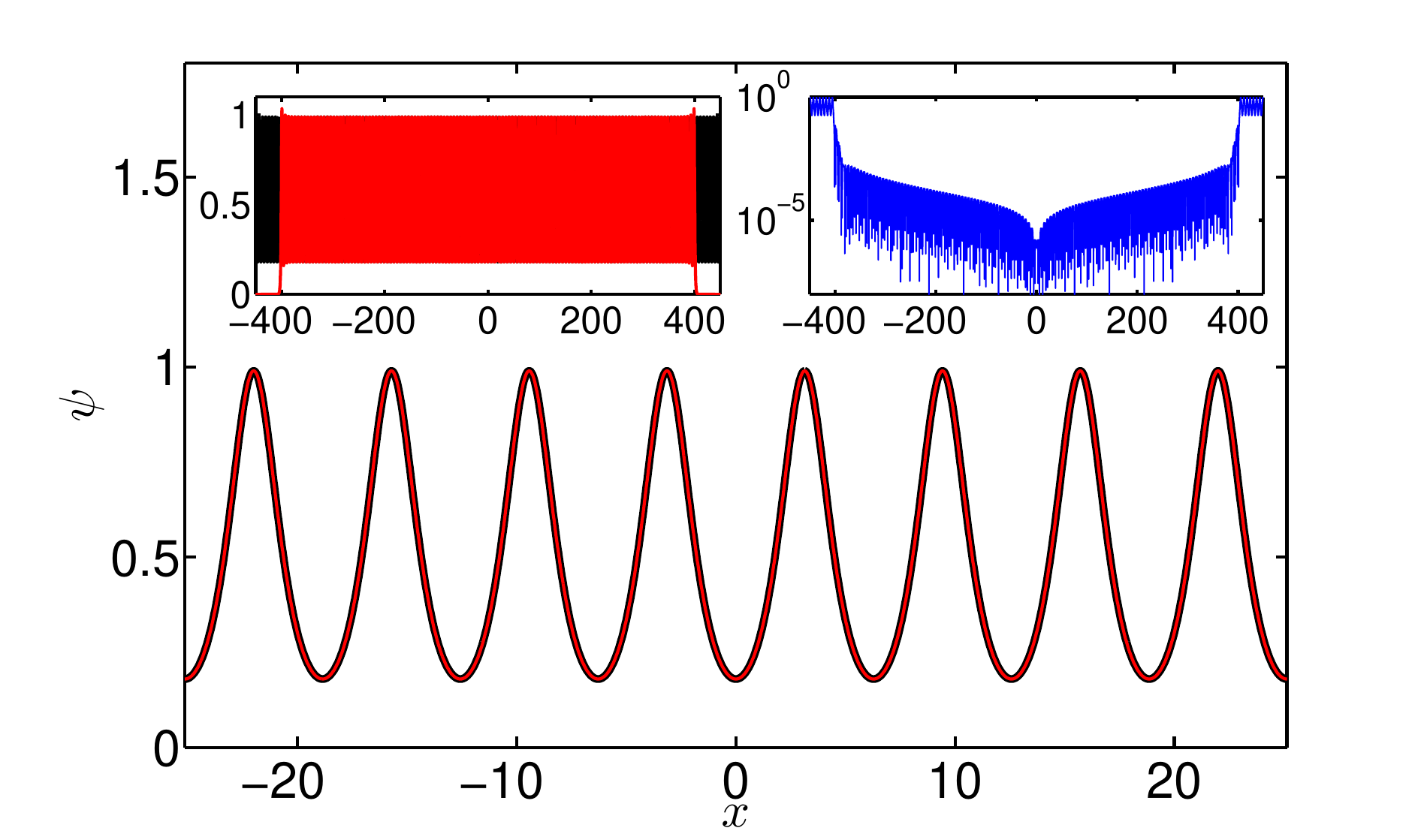}

	\caption{\small {\it (Color online)} 
		Cnoidal wave~\eqref{dn-branch} with real and imaginary half-periods $\omega_{0}=\pi$ and $\omega_{1}=1.6$ ($m\approx 0.48$, $\eta_1\approx 0.41$, $\eta_2\approx 0.59$) (thick black line) and its $128$-soliton model (thin red line) constructed from solitons having eigenvalues $\zeta_j \in (\eta_1, \eta_2)$, $j=1,...,128$, defined by~(\ref{eigen}) and norming constants~(\ref{DM-norming-constants-corrected}). 
		The left inset shows the same wave fields at larger scale, while the right inset illustrates the absolute difference between them (note the logarithmic vertical scale). 
		Note that both wave fields are real-valued and symmetric.
	}
	\label{fig:figCW}
\end{figure}

In the recent publication~\cite{gelash2021solitonic}, the expressions~(\ref{DM-norming-constants-corrected}),~(\ref{DM-positions-phases-box}) have been obtained analytically for the rectangular box-shaped potential, which contains a section of a plane wave in the middle and vanishes to zero at the edges. 
It has been done by removing the continuous spectrum from the scattering data and finding the specific corrections to the soliton norming constants due to the removal procedure, the resulting potential  being effectively equivalent to the numerical implementation of the genus zero soliton condensate with the Weyl DOS \eqref{weyl}. 
Here we use the expressions (\ref{DM-norming-constants-corrected}),~(\ref{DM-positions-phases-box}) as an ansatz, without solving the corresponding scattering problem. A similar ansatz was successfully used in \cite{congy_dispersive_2023} for the numerical implementation of KdV soliton condensates of different genera. Here the validity of this choice of the norming constants will be verified by the comparison of the numerically constructed $N$-soliton solution with the $\mathrm{dn}$ potential \eqref{dn-branch}.

Note that a multi-soliton solution having imaginary eigenvalues and norming constants~(\ref{DM-norming-constants-corrected}) is necessarily real-valued, $\psi_{N}(x)\in\mathbb{R}$, and symmetric, $\psi_{N}(x) = \psi_{N}(-x)$, see~\cite{gelash2021solitonic} for detail, that corresponds to the properties of cnoidal waves at $t=0$ discussed in Section~\ref{Sec:Sec2-1}.

Figure~\ref{fig:figCW} shows the comparison between the cnoidal wave \eqref{dn-branch} and its $128$-soliton approximation defined by ~(\ref{eigen}), (\ref{DM-norming-constants-corrected}). 
Here and below, unless otherwise indicated, we demonstrate our results on the example of a dn-branch cnoidal wave with real and imaginary half-periods $\omega_{0}=\pi$ and $\omega_{1}=1.6$ respectively ($m\approx 0.48$, $\eta_1\approx 0.41$, $\eta_2\approx 0.59$), as it enables us a direct comparison with the results published in~\cite{agafontsev2016integrable}. 
Note that we have checked cnoidal waves with other parameters and came to the same conclusions. 
As shown in the figure, the $128$-soliton solution ($128$-SS) model turns out to be practically indistinguishable from the corresponding cnoidal wave over almost its entire characteristic width $L_{N}$ found from the formula \footnote[1]{Formula \eqref{LN} comes from the definition of the density of states: indeed, $ U(\eta_2) = \int_{\eta_1}^{\eta_2} f(\eta) d\eta $ corresponds to the spatial density of solitons $N/L_N$.} 
\be \label{LN}
    L_{N} =\frac{N}{U(\eta_2)} = \frac{ N\pi}{\eta_2\bigg( E(1-\tilde{m}) - \big[1-\frac{E(\tilde{m})}{K(\tilde{m})}\big] K(1-\tilde{m}) \bigg)},
\ee
so that at $|x|\lesssim L_{N}/2$ the two solutions are practically indistinguishable from each other (the difference is of $10^{-4}$ order, see the right inset in Fig.~\ref{fig:figCW}) and at $|x|\gtrsim L_{N}/2$ the $128$-SS vanishes exponentially.


\begin{figure}[t]\centering
	\includegraphics[width=0.6\linewidth]{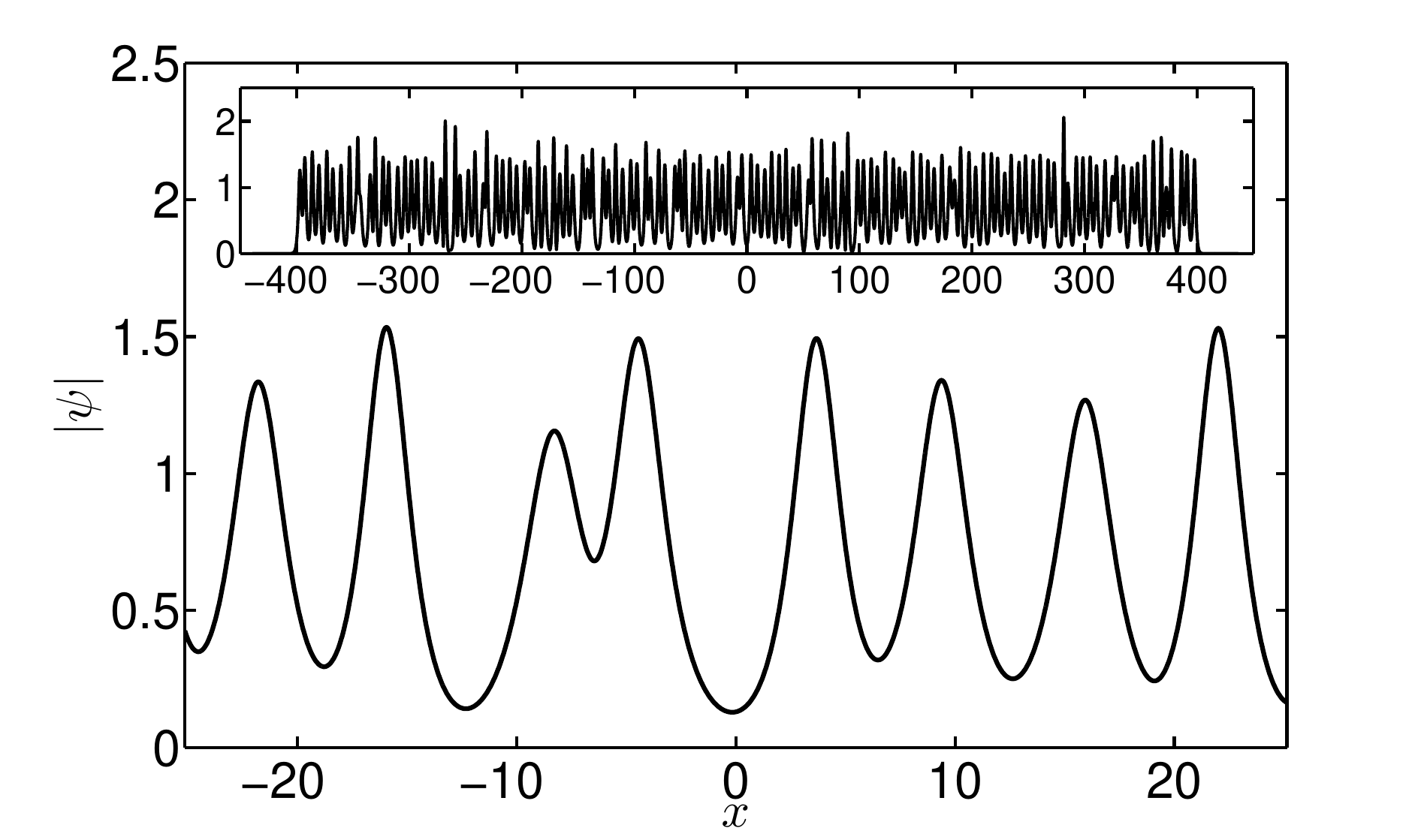}

	\caption{\small 
		Example of $128$-SS used for the modeling of the asymptotic statistically stationary state developing from the noise-induced MI of cnoidal wave with real and imaginary half-periods $\omega_{0}=\pi$ and $\omega_{1}=1.6$. 
		The $128$-SS is constructed from solitons having eigenvalues $\zeta_n$~(\ref{eigen}), and random soliton positions and phases uniformly distributed over the intervals $x_{0n}\in [-2,2]$ and $\theta_{0n}\in [0,2\pi)$. 
		The inset shows the same wave field on a larger scale.
	}
	\label{fig:figCW-SG}
\end{figure}

\begin{figure}[t]\centering
	\includegraphics[width=0.8\linewidth]{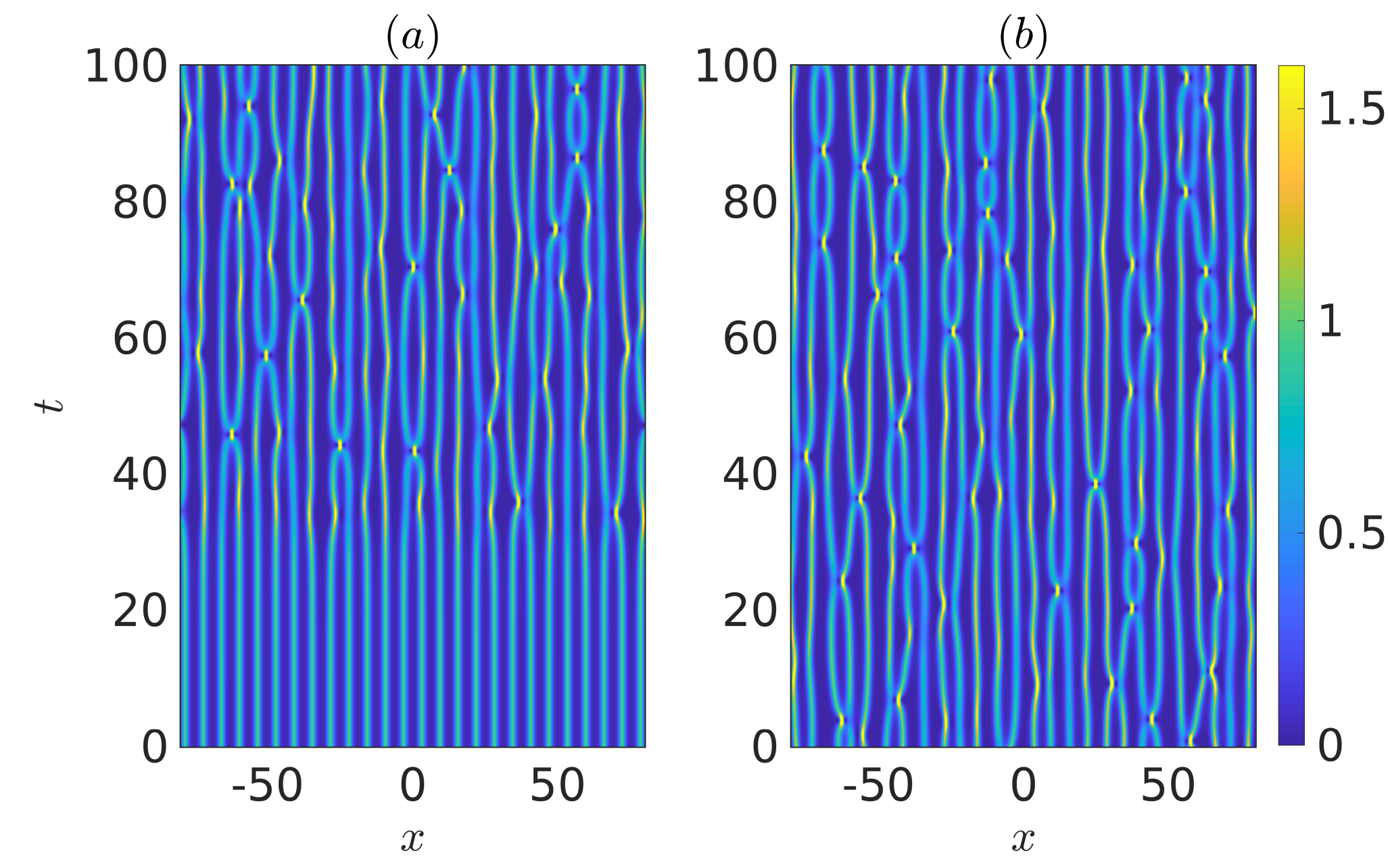}

	\caption{\small {\it (Color online)} MI vs. SG---numerical simulations of the fNLSE: Space-time diagrams of $|\psi(x,t)|^2$. (a) Noise-induced MI of a cnoidal wave; (b) Random phase bound-state $N$-SS  approximating genus 1 soliton condensate
[the initial amplitude $|\psi (x,0)|$ is shown in Fig.~\ref{fig:figCW-SG}].
	}
	\label{fig:evol}
\end{figure}

\begin{figure*}[t]\centering
	\includegraphics[width=0.49\linewidth]{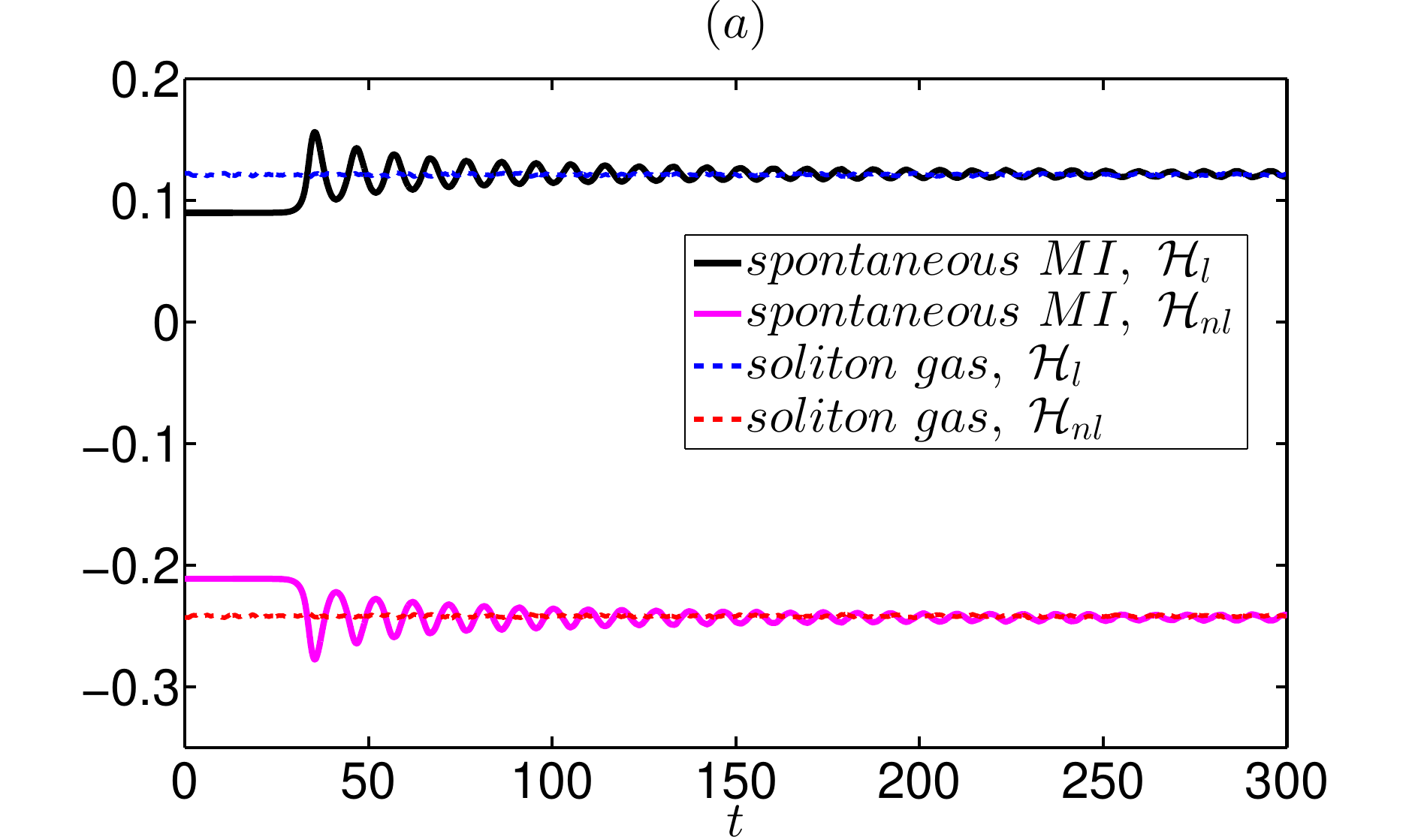}
	\includegraphics[width=0.49\linewidth]{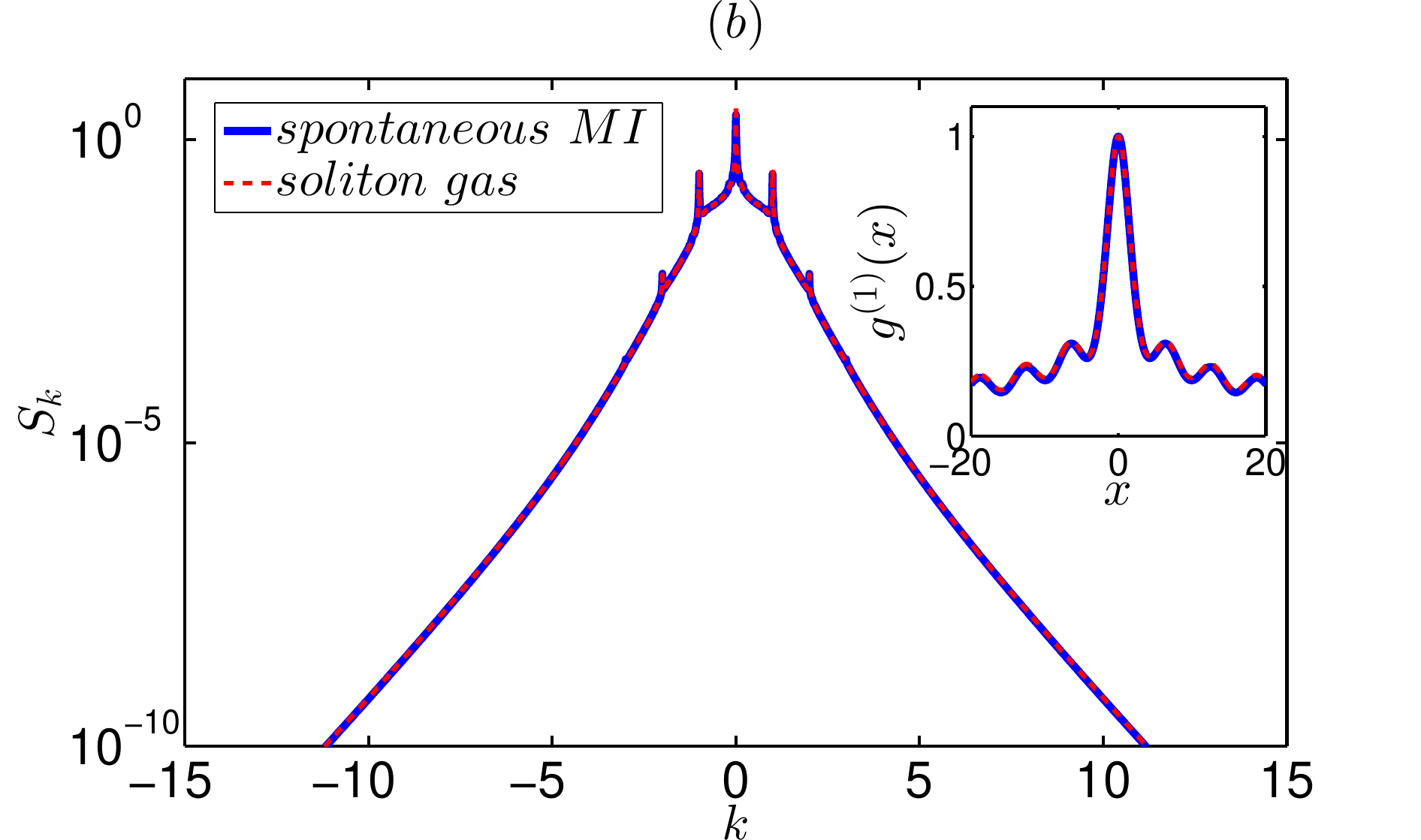}\\[5mm]
	\includegraphics[width=0.49\linewidth]{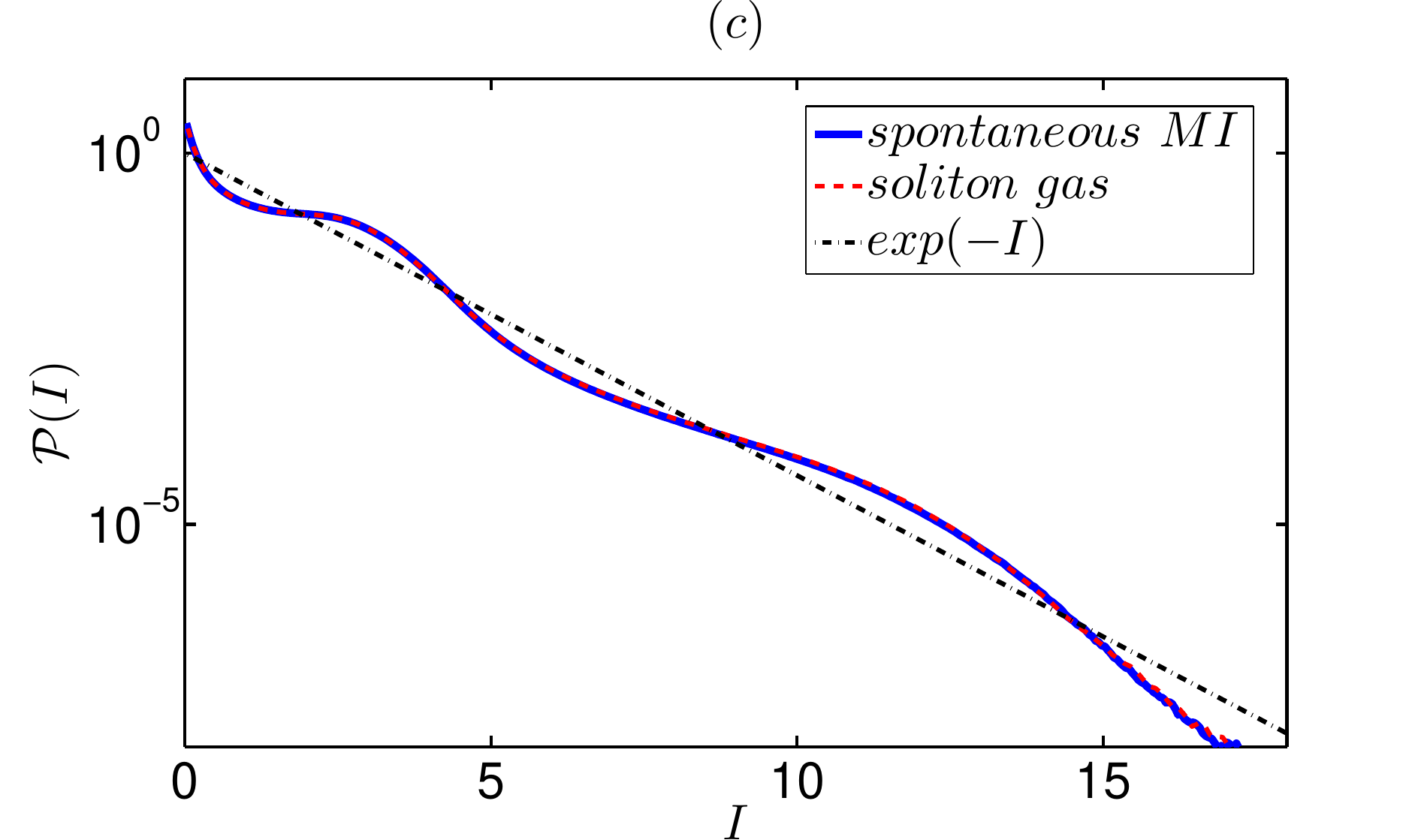}
	\includegraphics[width=0.49\linewidth]{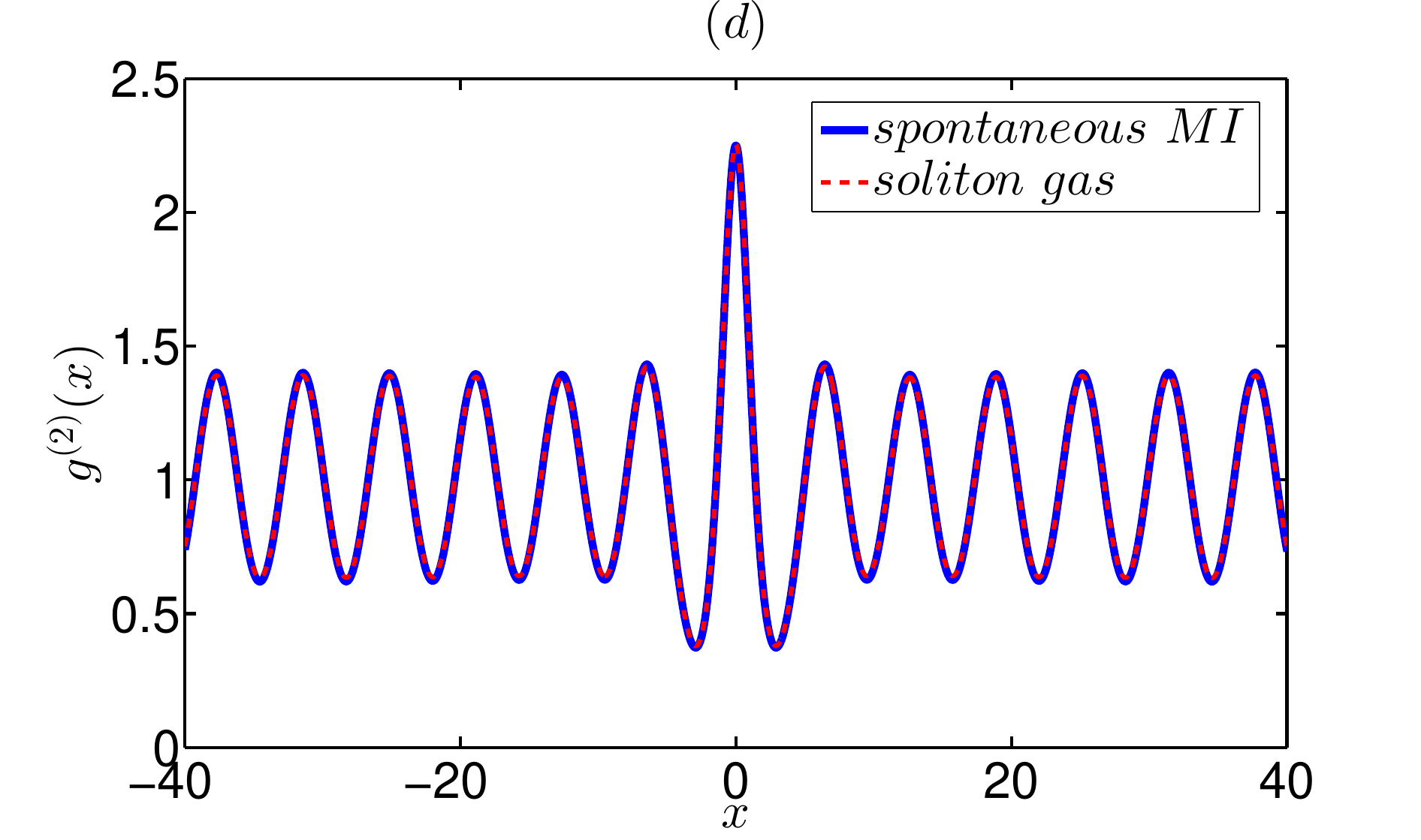}

	\caption{\small {\it (Color online)} 
		Comparison of the ensemble-averaged statistical characteristics between the spontaneous (noise-induced) MI and the random-phase $128$-SS configured according to the genus 1 soliton condensate DOS \eqref{DOS_cond} with $\eta_1 \approx 0.31, \eta_2 \approx 0.59, m \approx 0.48$: 	
		(a) time evolution of the kinetic $\langle\mathcal{H}_{l}\rangle$ and potential $\langle\mathcal{H}_{nl}\rangle$ energies, 
		(b) the Fourier spectrum $S_{k}$, (c) the PDF $\mathcal{P}(I)$ of relative wave intensity $I = |\psi|^{2}/\langle\overline{|\psi|^{2}}\rangle$ and (d) the autocorrelation of intensity $g^{(2)}(x)$. 
		Panels (b,c,d) compare the statistical functions between the asymptotic statistically stationary state of the spontaneous MI and the random-phase $128$-SS; the statistical functions here are additionally averaged over time intervals $t\in[240,300]$ for the MI and $t\in[0,300]$ for the $128$-SS. 
		The black dash-dot line in panel (c) indicates the exponential distribution $\mathcal{P}_{e}(I)=e^{-I}$, while the inset in panel (b) shows the correlation function $g^{(1)}(x)$. 
	}
	\label{fig:figStat}
\end{figure*}

\subsection{Noise-induced MI vs. soliton condensate: statistical properties}
\label{Sec:Sec3-2}

In this subsection, we quantitatively compare the statistical properties of an asymptotic statistically stationary state developing from the noise-induced MI of a cnoidal ($\mathrm{dn}$) wave and of the SG---the genus 1 bound state soliton condensate modeling this asymptotic state.

The dense SG sufficiently close to a condensate is built as a random ensemble containing $200$ realizations of $N$-SS with $N=128$, the soliton eigenvalues prescribed by~\eqref{phi},~\eqref{sample}, 
and random soliton phases uniformly distributed over the interval $\theta_{0n}\in [0,2\pi)$.  
Note that, when all solitons sit at the origin, $x_{0n}=0$, the corresponding multi-soliton solution is symmetric in space. 
Following~\cite{gelash2019bound}, we use soliton positions distributed over some sufficiently  small interval, e.g.,  $x_{0n}\in [-2,2]$, as it allows us to avoid this artificial symmetry and does not change significantly the spatial density of solitons, cf. Figs.~\ref{fig:figCW} and~\ref{fig:figCW-SG}. 
The SG constructed with this procedure is the so-called {\it diluted soliton condensate} \cite{congy_dispersive_2023} and it is described by the DOS $f(\eta)=Cf^{(1)}(\eta;\eta_1,\eta_2)$ where $C$ is the dilution parameter, which in our case is very close to unity, $C\approx 0.9997$; see Appendix \ref{Sec:Implementation} for details. 
An example of one such $128$-SS is shown in Fig.~\ref{fig:figCW-SG}; as in the previous subsection, its wave field remains of unity order at $|x|\lesssim L_{N}/2$ and becomes small at $|x|\gtrsim L_{N}/2$.
Also note that in our SG, all realizations of the $N$-SS have exactly the same set of soliton eigenvalues \eqref{phi}, \eqref{sample}. 
We have tested SGs with eigenvalues randomly distributed according to the density function \eqref{phi} and came to practically the same results.  

In Fig.~\ref{fig:evol} we present two contour plots of intensity $|\psi(x,t)|^{2}$ showing the space-time evolution of the spontaneous MI of a cnoidal wave (a), and of the SG modeled by the random $N$-SS (b). 
One  can see that, at sufficiently large time $t$, the MI displays strikingly similar random wave patters to the SG case which we quantitatively analyze and compare below.

The long-term evolution of the noise-induced MI can be characterized by the stationary values of the kinetic $\langle\mathcal{H}_{l}\rangle$ and potential $\langle\mathcal{H}_{nl}\rangle$ energies, and also by the stationary shapes of the Fourier spectrum $S_{k}\propto \langle|\psi_{k}|^{2}\rangle$, the probability density function $\mathcal{P}(I)$ of the relative wave intensity $I=|\psi|^{2}/\langle\overline{|\psi|^{2}}\rangle$, and the autocorrelation function of the intensity $g^{(2)}(x)$, see e.g.~\cite{agafontsev2015integrable,agafontsev2016integrable,kraych2019statistical} and Appendix~\ref{Sec:NumMethods}. 
Here $\langle ...\rangle$ denotes averaging over the ensemble of initial conditions (e.g., random realizations of the initial noise), while the overline denotes spatial averaging over the simulation box $x\in[-L/2, L/2]$. Due to the periodic boundary conditions the average intensity
\begin{eqnarray}
	\overline{|\psi|^{2}} = \frac{1}{L}\int_{-L/2}^{L/2}|\psi|^{2}\,dx = \mathcal{N} \label{wave-action-MT}
\end{eqnarray}
is the integral of motion, one among the infinite number of conserved quantities of the fNLSE evolution~\cite{novikov1984theory}. 
The kinetic energy $\mathcal{H}_{l}$ (related to dispersion) and potential energy $\mathcal{H}_{nl}$ (related to nonlinearity) are the two parts of the total energy (Hamiltonian) $\mathcal{E}$, which is another integral of motion, 
\begin{eqnarray}
	\mathcal{E} = \mathcal{H}_{l} + \mathcal{H}_{nl}, \quad \mathcal{H}_{l} = \frac{1}{L}\int_{-L/2}^{L/2}|\psi_{x}|^{2}\,dx,\quad
	\mathcal{H}_{nl} = -\frac{1}{L}\int_{-L/2}^{L/2}|\psi|^{4}\,dx. \label{energy-MT}
\end{eqnarray}
Following~\cite{agafontsev2015integrable}, we compute statistical functions for the noise-induced MI by simulating the time evolution within the fNLSE~(\ref{NLSE}) starting from $1000$ superpositions of cnoidal wave with random noise, see Appendix~\ref{Sec:NumMethods} for detail; then, we average the results over realizations of initial noise. 
For the soliton gas case, we average the results over $200$ realizations of $N$-SS. 

As shown in Fig.~\ref{fig:figStat}(a), for the noise-induced MI, the kinetic and potential energies approach  the stationary values of $\langle\mathcal{H}_{l}\rangle\approx 0.12$ and $\langle\mathcal{H}_{nl}\rangle\approx -0.24$ during the oscillatory transient evolution. 
For the SG case, the kinetic and potential energies have the same values from the start and do not change during the evolution in time. 

Then, we compare the statistical functions for the long-time statistically stationary state of the noise-induced MI and for the SG cases. 
To improve accuracy in the calculation of these functions, we perform ensemble-averaging together with the temporal averaging. 
For the MI, the latter is done over the interval $t\in[240,300]$, which is sufficiently close to the asymptotic stationary state, while for the SG---from the start of the evolution, $t\in[0,300]$. 
Also note that, to avoid edge effects, for the SG case all the statistical functions except the Fourier spectrum $S_{k}$ are computed within the central part $x\in[-\ell,\ell]$, $\ell=250$, of the $128$-SS, while $S_{k}$ is renormalized proportionally to the spatial extent of the wave field; see Appendix~\ref{Sec:NumMethods} for detail. 

As shown in Fig.~\ref{fig:figStat}(b-d), the Fourier spectrum $S_{k}$, the PDF $\mathcal{P}(I)$ and the autocorrelation function $g^{(2)}(x)$ turn out to be nearly identical for the SG  and for the asymptotic statistically stationary state of the noise-induced MI. 
To additionally verify our results on the Fourier spectrum, we consider the spatial correlation function $g^{(1)}(x)$, which in the periodic case can also be calculated via the inverse Fourier transform of the spectrum, $g^{(1)}(x) \propto\mathcal{F}^{-1}[S_{k}]$. 
We compute this correlation function directly according to its definition given in Appendix~\ref{Sec:NumMethods} (for the SG case this is done within the central part of the wave field $x\in[-\ell, \ell]$, $\ell=250$), and observe practically identical results for the SG and for the noise-induced MI; see the inset in Fig.~\ref{fig:figStat}(b). 
The latter confirms the coincidence of the Fourier spectra for the noise-induced MI and SG-- without any renormalization of the correlation functions. 

We conclude that (i) the asymptotic statistically stationary state developing from the noise-induced MI of a cnoidal wave, and (ii) the constructed SG (the genus 1 soliton condensate realized via $N$-SS) are characterized by the identical statistical functions. 


\begin{figure*}[t]\centering
	\includegraphics[width=0.49\linewidth]{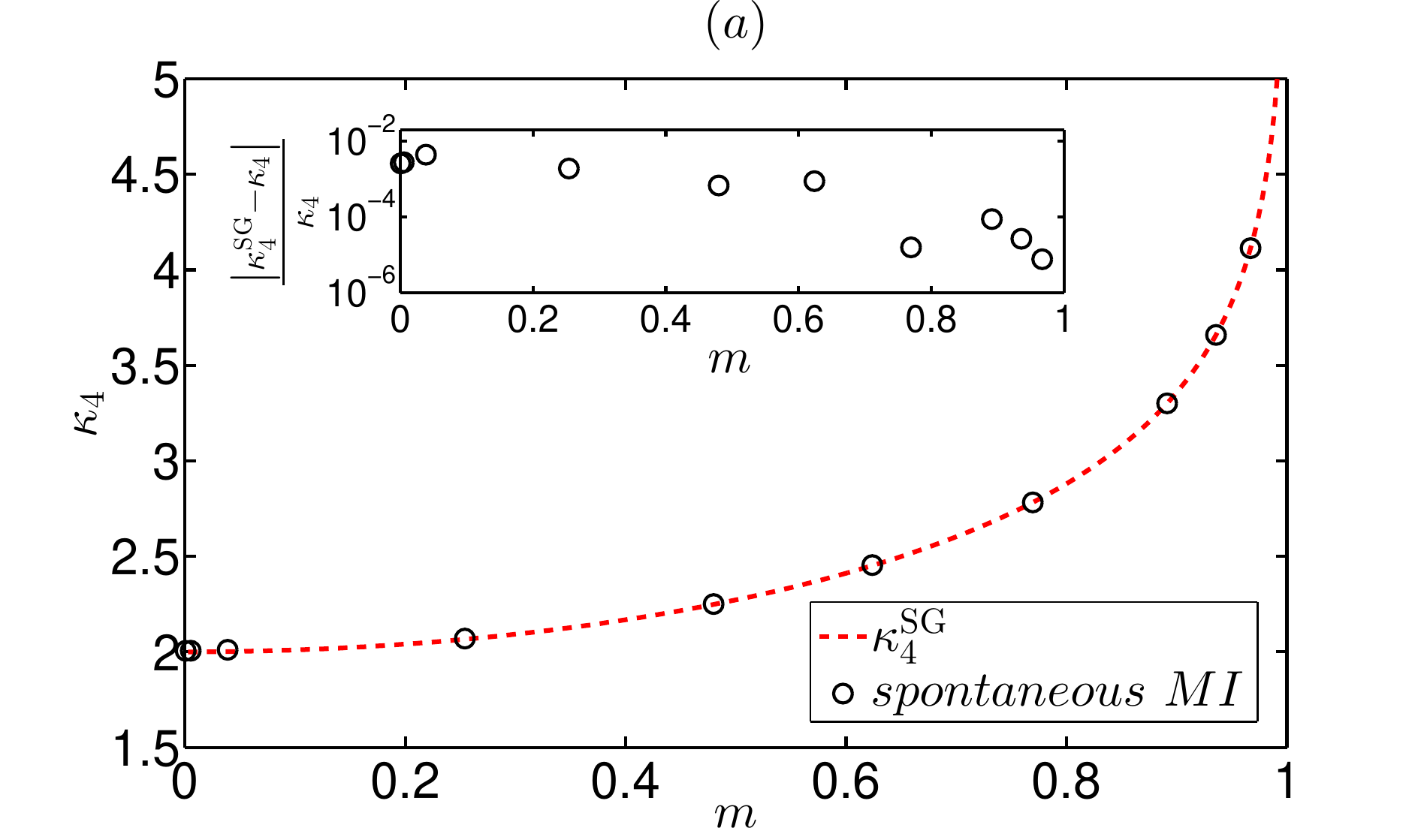}
	\includegraphics[width=0.49\linewidth]{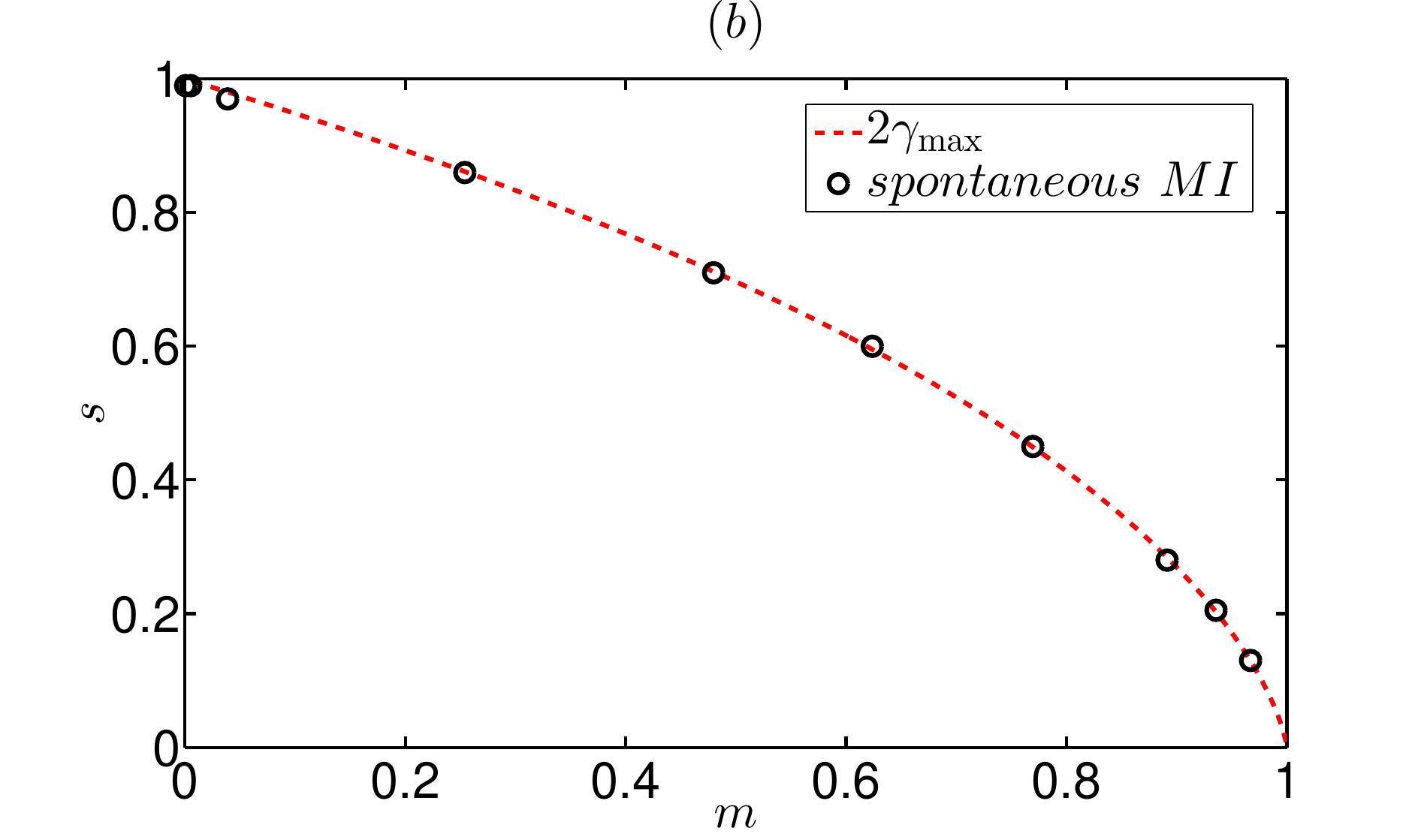}

	\caption{\small {\it (Color on-line)} 
	    The dependence (black circles) on the elliptic modulus $m$ of (a) the kurtosis $\kappa_{4}$ in the statistically stationary state of the MI and (b) the frequency $s$ of its temporal oscillations in the nonlinear stage of the MI. 
		In panel (a), the dashed red line indicates the kurtosis $\kappa_{4}^{\mathrm{SG}}$~(\ref{k4-SG}) for the soliton gas case, while the inset shows the relative difference $|\kappa_{4}^{\mathrm{SG}}-\kappa_{4}|/\kappa_{4}$ (note the logarithmic vertical scale). 
		In panel (b), the dashed red line illustrates the frequency $s^{\mathrm{SG}} = 2\gamma_{\max}$ obtained in the soliton gas model approximation, see~(\ref{theta-evolution})-(\ref{s-SG}).
	}
	\label{fig:figW1vsK4}
\end{figure*}

\subsection{Nonlinear stage of the  MI development  and the asymptotic value of the kurtosis}
\label{Sec:Sec3-3}

As we have seen in the previous subsection, the SG approximation can be used to accurately model the statistical properties of the noise-induced MI. 
Here we apply this approximation to study the properties of the kurtosis $\kappa_{4} = \langle|\psi|^{4}\rangle/\langle|\psi|^{2}\rangle^{2}$ in the nonlinear stage of MI.

First, for a bound-state SG with the condensate DOS $f^{(1)}(\eta;\eta_1, \eta_2)$ \eqref{DOS_cond}, one can use formulas (\ref{moments24}) and get the following expression for the kurtosis,
\begin{eqnarray}
	\kappa_{4}^{\mathrm{SG}} = \frac{\langle|\psi|^{4}\rangle}{\langle|\psi|^{2}\rangle^{2}} = 2\frac{\eta_1^4+\frac23 \eta_2^2 \eta_1^2+ \eta_2^4 - \frac{4}{3} (\eta_1^2+\eta_2^2)w^2}{\left[\eta_1^2+\eta_2^2 - 2w^2\right]^{2}}. \label{k4-SG}
\end{eqnarray}
Evaluating the stationary values of $\kappa_{4}$ for the MI of cnoidal waves with $m \in [0.0007, 0.97]$ we observe a very good agreement with~(\ref{k4-SG}): compare the black circles and the dashed red line in Fig.~\ref{fig:figW1vsK4}(a). 
The difference between the computed values $\kappa_{4}$ for the MI and $\kappa_{4}^{\mathrm{SG}}$ is a small positive number within $0.5$\%  of $\kappa_{4}$ for cnoidal waves with $m\leq 0.62$, while for larger $m$ it alternates in sign and decreases below $10^{-4}$ order, see the inset in Fig.~\ref{fig:figW1vsK4}(a). 

Let us now consider the temporal oscillations of the kurtosis, which are observed in the beginning of the nonlinear stage of MI, see Fig.~\ref{fig:figStat} (a).
For the first time, these oscillations have been described in~\cite{agafontsev2015integrable} for the noise-induced MI developing from the plane wave~(\ref{pw}), and they affect not only the kurtosis, but all the other statistical functions as well. 
For the kurtosis, which is linearly connected with the potential energy as $\kappa_{4} = -\langle\mathcal{H}_{nl}\rangle\,/\langle\mathcal{N}\rangle^{2}$, see~(\ref{wave-action-MT})-(\ref{k4-SG}), these oscillations are very well approximated by the following ansatz, 
\begin{eqnarray}
	\kappa_{4}(t) \approx \kappa_{4}^{\mathrm{A}} + \frac{p}{t^{3/2}}\,\sin\left(s\,t + \frac{q}{\sqrt{t}} + \phi_{0}\right).
	\label{oscillations-ansatz}
\end{eqnarray}
Here $s$ is the frequency, $\phi_{0}$ is the initial phase, $\kappa_{4}^{\mathrm{A}}$ is the asymptotic stationary value of $\kappa_{4}$, while the terms $p/t^{3/2}$ and $q/\sqrt{t}$ describe the decay in amplitude and the nonlinear phase shift respectively. 
Note that ansatz~(\ref{oscillations-ansatz}) is applicable over sufficiently long time intervals: for instance, in Fig.~9 of~\cite{agafontsev2015integrable} it approximates very well more than $60$ oscillation periods using the same set of parameters $s$, $p$, $q$ and $\phi_{0}$. 
The numerically computed oscillation frequency has been found to be very close to the double maximum growth rate of the MI, $s\approx 2\gamma_{\max}$. 
Note that the frequency $s$ represents the characteristic of the \textit{nonlinear} stage of MI, while the maximum growth rate $\gamma_{\max}$ characterizes the \textit{linear} stage of the MI development. 

In~\cite{agafontsev2016integrable}, the very similar oscillations with amplitude decaying according to a slightly different power law $\propto t^{-\alpha}$, $1 < \alpha < 1.5$, have been observed for the noise-induced MI developing from cnoidal waves. 
The relation
\begin{eqnarray}
	s = 2\gamma_{\max}
	\label{s-gamma-max}
\end{eqnarray}
has been confirmed numerically for all examined cnoidal waves with $m \in [0.0007,  0.97]$). 
In the present paper, the oscillations of this type for the kinetic and potential energies can be seen in Fig.~\ref{fig:figStat}(a). 

The exact multi-soliton solutions employed in the previous subsections for the numerical implementation of the soliton condensate can be interpreted as approximations of (i) the unperturbed cnoidal wave and (ii) the asymptotic statistically stationary state of the noise-induced MI developed from this wave. 
The main difference between these two solitonic approximations is that the soliton phases $\theta_{0n}$, see~(\ref{Ck_param}), are strongly correlated~(\ref{DM-positions-phases-box}) in the first case and are random in the second case.  Given the isospectrality of the integrable fNLSE evolution it is then reasonable to assume that the  development of integrable turbulence occurs due to the gradual desynchronization of soliton phases during the  evolution in time. 
Indeed, the soliton norming constants evolve as 
\begin{eqnarray}
	C_{n}(t) = C_{n}(0)\,e^{-4i\lambda_{n}^{2}t}, 
	\label{C-evolution}
\end{eqnarray}
so that, for imaginary eigenvalues $\lambda_{n}=i\eta_{n}$, the soliton positions do not change, see~(\ref{Ck_param}), while the phases rotate as
\begin{eqnarray}
	\theta_{n}(t) = \theta_{n0} + 4\eta_{n}^{2}\,t. 
	\label{theta-evolution}
\end{eqnarray}
In the general case, the rotation frequencies $4\eta_{n}^{2}$ are incommensurable, and the gradual mismatch in soliton phases is at the core of the MI evolution towards the long-time statistically stationary state. 
In this interpretation, the oscillations of the statistical functions are observed when the correlation of the phases is still present, but is vanishing during the evolution in time. 

In the soliton condensate with the DOS \eqref{DOS_cond}, a significant number of solitons  have eigenvalues sufficiently close to either $i\eta_{1}$, or $i\eta_{2}$, see Fig.~\ref{fig:genus0}(b), with the phase rotation frequencies close to $4\eta_{1}^{2}$ and $4\eta_{2}^{2}$, respectively. 
For instance, in the example considered here, approximately 41\%
of  solitons have a parameter $\zeta_n \in [\eta_1,\eta_1+ \varepsilon]\cup [\eta_2-\varepsilon,\eta_2]$ for $\varepsilon = 0.1 (\eta_2-\eta_1)$. 
It is then natural to assume, at least at the phenomenological level, that these two groups of solitons would have a dominant effect on the statistical parameters observed in the development of MI. 
Then, if at time $t$ the soliton groups concentrated near $i\eta_{1}$ and at $i\eta_{2}$ have a specific joint set of phases, then at $t+\Delta T$, where $\Delta T$ is defined by 
\begin{eqnarray}
	(4\eta_{2}^{2} - 4\eta_{1}^{2})\Delta T = 2\pi,
	\label{delta-T}
\end{eqnarray}
they will have practically the same set of phases, since the first group will have its phase rotated by $4\eta_{1}^{2}\Delta T$ and the second group -- by $4\eta_{2}^{2}\Delta T = 4\eta_{1}^{2}\Delta T + 2\pi$. 
The rotation of all phases by the same angle $4\eta_{1}^{2}\Delta T$ is not important, as it corresponds to the change in phase of the whole wave field $\psi\to \psi\,e^{4i\eta_{1}^{2}\Delta T}$, which is ignored in the considered statistical functions. 
The described above behavior corresponds to oscillations with frequency
\begin{eqnarray}
	s^{\mathrm{SG}} = \frac{2\pi}{\Delta T} = 4\left[\eta_{2}^{2} - \eta_{1}^{2}\right],
	\label{s-SG}
\end{eqnarray}
which is nothing but the ``breathing'' frequency associated with the bound-state two-soliton fNLSE solution with the eigenvalues $i\eta_1$ and $i\eta_2$~\cite{novikov_theory_1984}. One can  see that the frequency \eqref{s-SG} 
is exactly twice as large  as the maximum growth rate $\gamma_{\max}$ of  the MI, see~(\ref{gamma_max}), 
\begin{eqnarray}
	s^{\mathrm{SG}} = 2\gamma_{\max}, 
	\label{s-SG-gamma-max}
\end{eqnarray}
which coincides with the relation~(\ref{s-gamma-max}) observed in direct numerical simulations of \cite{agafontsev2016integrable}. This agreement supports our assumption about the dominant role of the two groups of solitons with the eigenvalues close to $i\eta_1$ and $i \eta_2$ and, more generally, provides further confirmation of the validity of the soliton condensate model of the MI.


\section{Conclusions}
\label{Sec:Conclusions}

In this paper we have developed a soliton gas model of the  integrable turbulence resulting from the long-time development of the spontaneous (noise-induced) modulational instability of the elliptic  $\mathrm{dn}$ family of nonlinear periodic solutions of the fNLSE. This fundamental type of integrable turbulence was investigated numerically in \cite{agafontsev2016integrable}  with a number of peculiar properties observed. 
The key role in our construction is played by the spectral theory of the critically dense soliton gases, the so-called  soliton condensates \cite{el_spectral_2020}. Building upon the recently developed solitonic model of the spontaneous modulational instability of the plane wave solutions \cite{gelash2019bound}  we have shown that the bound state (non-propagating) genus one soliton condensate, whose spectral support coincides with the Zakharov-Shabat finite-gap spectrum of the $\mathrm{dn}$ solution, provides a highly accurate model of the long-time development of the ``elliptic'' integrable turbulence.

We have used  the  spectral theory of fNLSE soliton gases \cite{el_spectral_2020, congy_statistics_2024} to obtain analytical expressions for some observables (ensemble averages) in the soliton condensate approximating integrable turbulence. 
This enabled us to evaluate the kurtosis of integrable turbulence as a function of the elliptic parameter $m$ of the initial $\mathrm{dn}$ data.
We then constructed the soliton condensate numerically by building the $N$-soliton ensemble, $N$ large, configured according to the spectral density of states \eqref{DOS_cond} of the genus 1 condensate complemented by a special choice of the soliton norming constants, and compared the statistical functions of the ``numerical'' soliton condensate with those  of the integrable turbulence realized in direct fNLSE numerical simulations. In all cases an excellent agreement was observed. The soliton condensate model also enabled us to quantify the peculiar  oscillations of the kurtosis observed
at the intermediate stage of the integrable turbulence development.

Our results confirm the validity of the soliton gas model of the spontaneous MI of nonlinear periodic waves in self-focusing media and also provide a strong indication that a similar modeling   can be applied to  a class of quasiperiodic finite-gap potentials using  appropriate soliton condensates of higher genera. More generally, our results provide further evidence of the efficacy of the  soliton gas framework for the modeling complex emergent phenomena in nonlinear dispersive waves.

\medskip
\begin{center}
\textbf{Acknowledgements}
\end{center}
We thank the Isaac Newton Institute for Mathematical Sciences, Cambridge, for support and hospitality during the programme Emergent Phenomena in Nonlinear Dispersive Waves, where the work on this paper was partially undertaken. The work of G.E., T.C. and G.R. at the INI was supported by EPSRC grant EP/V521929/1. 
This work has been partially supported by the Agence Nationale de la Recherche through the LABEX CEMPI project (ANR-11-LABX-0007), the SOGOOD project (SOGOOD ANR-21-CE30-0061), the Ministry of Higher Education and Research, Hauts de France council and European Regional Development Fund (ERDF) through the Nord-Pas de Calais Regional Research Council and the European Regional Development Fund (ERDF) through the Contrat de Projets Etat-Région (CPER Wavetech). 
P.S. and S.R. thank the Centre d’Etudes et de Recherches Lasers et Applications CERLA for the technical help and equipment. 
The work of D.A. was supported by the Russian Science Foundation (Grant 19-72-30028). 
D.A. also wishes to thank the Isaac Newton Institute and London Mathematical Society for the financial support on the Solidarity Programme, as well as the Department of Mathematics, Physics and Electrical Engineering at Northumbria University for hospitality.


%

\newpage

\appendix

\section*{Appendix}

\medskip
\section{Numerical methods}
\label{Sec:NumMethods}

To study the properties of the asymptotic statistically stationary state developing from the noise-induced MI of a cnoidal wave, we first need to reach this state. 
For this purpose, following~\cite{agafontsev2016integrable}, we solve~(\ref{NLSE}) numerically starting from a superposition of cnoidal wave~(\ref{dn-branch}) and random, statistically homogeneous in space noise, 
\begin{eqnarray}
	\psi|_{t=0} &=& \psi_{\mathrm{dn}}(x,0) + \epsilon(x), \nonumber\\
	\epsilon(x) &=& A_{0} \sqrt{\frac{G_{n}}{\theta L}} \sum_{m} e^{-|k_{m}|^{n}/\theta^{n} + i\phi_{m} + ik_{m}x}. \label{initial-conditions}
\end{eqnarray}
Here $A_{0}$ is the noise amplitude, $\theta$ is noise spectral width, $k_{m}=2\pi\, m/L$ is the wavenumber, $m\in\mathbb{Z}$ is integer, $\phi_{m}\in [0,2\pi)$ are random phases for each $k_{m}$ and each noise realization, $n\in\mathbb{N}$ is the exponent defining the shape of noise spectrum, $G_{n} = \pi\,2^{1/n}/\Gamma_{1+1/n}$ is normalization constant such that the average noise intensity $\langle\overline{|\epsilon(x)|^{2}}\rangle$ equals $A_{0}^{2}$, and $\Gamma$ is Euler's gamma function. 
We use parameters $n=32$, $A_{0}=10^{-5}$ and $\theta=5$, which are slightly different compared to~\cite{agafontsev2016integrable} where $n=2$ has been used; this leads us to slightly different timing when the MI enters its nonlinear stage, but the other results coincide with those reported in that paper.

For the numerical modeling of~(\ref{NLSE}), we use the pseudo-spectral Runge-Kutta fourth-order method in adaptive grid with the grid size $\Delta x$ set from analysis of the Fourier spectrum of the solution, see~\cite{agafontsev2015integrable} for detail. 
The simulation box $x\in [-L/2, L/2]$ has periodic boundaries. 
For the cnoidal wave with real and imaginary half-periods $\omega_{0}=\pi$ and $\omega_{1}=1.6$, we use $L=256\pi$ and start simulations on the grid of $16\,384$ nodes, reaching the final simulation time $t_{f}=300$ when the statistical functions are practically stationary, see Fig.~\ref{fig:figStat}(a). 
Note that the number of nodes changes adaptively during the simulations between $16\,384$ and $262\,144$, and for other cnoidal waves we sometimes have to use different parameters; see~\cite{agafontsev2016integrable} for detail. 
For each of the studied cnoidal wave, we simulate the time evolution for $1000$ random realizations of the initial conditions~(\ref{initial-conditions}) and then average the results over these realizations. 
To improve the accuracy in the measurement of the stationary values of the statistical functions and exclude influence of the residual temporal oscillations, see e.g.~\cite{agafontsev2015integrable,agafontsev2016integrable} and Fig.~\ref{fig:figStat}(a), we perform an additional averaging over time interval placed sufficiently far in the nonlinear stage of the MI; for cnoidal wave with $\omega_{0}=\pi$ and $\omega_{1}=1.6$ ($m\approx 0.48$), this interval is $t\in [240, 300]$. 

We model the SG as a random ensemble containing $200$ realizations of $N$-SS with $N=128$, and the soliton eigenvalues and norming constants chosen as described in Section~\ref{Sec:Sec3-2}. 
Computation of the wave fields is performed in the box $x\in [-\tilde{L}/2, \tilde{L}/2]$, $\tilde{L} = 384\pi$, which contains $65\,536$ nodes, by using the dressing method~\cite{novikov1984theory,zakharov1978relativistically} combined with $100$-digits precision arithmetics; see~\cite{gelash2018strongly,gelash2019bound} for detail. 
As shown in Fig.~\ref{fig:figCW-SG}, the constructed $128$-SS turn out to be of unity order within a smaller interval $x\in[-L_{N}/2, L_{N}/2]$, $L_{N}\simeq 256\pi$, and decay with increasing $|x|$ outside this interval. 
The decay is exponential, so that at the edges of the computational box $|x|\simeq \tilde{L}/2$ these wave fields are of $10^{-20}$ order or smaller. 
The latter allows us to simulate the time evolution within~(\ref{NLSE}) starting from these $128$-SS by using the same numerical scheme with periodic boundary conditions as described above for the MI case. 
Doing so, we observe that the corresponding statistical functions, averaged over the ensemble of $200$ realizations, do not change with time; e.g., see the dashed lines in Fig.~\ref{fig:figStat}(a) for the evolution of kinetic and potential energies. 
Hence, the constructed soliton gas already rests in the statistically stationary state, and to improve the accuracy in the computation of its statistical properties, we perform an additional averaging over the time interval $t\in [0, 300]$.

The fNLSE~(\ref{NLSE}) conserves an infinite series of invariants~\cite{novikov1984theory}, which can be written in the form 
\begin{eqnarray}
	\mathcal{I}_{j} &=& \frac{1}{L}\int_{-L/2}^{L/2}\psi\,\mathcal{A}_{j}\,dx, \label{integrals_rec1}\\
	\mathcal{A}_{j} &=& \frac{\partial\mathcal{A}_{j-1}}{\partial x} + \psi\sum_{l+m=j-1}\mathcal{A}_{l}\mathcal{A}_{m}, \label{integrals_rec2}
\end{eqnarray}
where $\mathcal{A}_{1}=\psi^{*}$. 
The first three invariants are the wave action (the average intensity of fNLSE wave field),
\begin{equation}\label{wave-action}
	\mathcal{N} = \overline{|\psi|^{2}} = \frac{1}{L}\int_{-L/2}^{L/2}|\psi|^{2}\,dx = \sum_{k}|\psi_{k}|^{2},
\end{equation}
the momentum
\begin{equation}\label{momentum}
	\mathcal{M} = \frac{i}{2L}\int_{-L/2}^{L/2}(\psi_{x}^{*}\psi-\psi_{x}\psi^{*})\,dx = \sum_{k}k|\psi_{k}|^{2},
\end{equation}
and the total energy
\begin{eqnarray}
	&& \mathcal{E} = \mathcal{H}_{l} + \mathcal{H}_{nl}, \label{energy-1}\\
	&& \mathcal{H}_{l} = \overline{|\psi_{x}|^{2}} = \frac{1}{L}\int_{-L/2}^{L/2}|\psi_{x}|^{2}\,dx = \sum_{k}k^{2}|\psi_{k}|^{2}, \label{energy-2}\\
	&& \mathcal{H}_{nl} = -\overline{|\psi|^{4}} = -\frac{1}{L}\int_{-L/2}^{L/2}|\psi|^{4}\,dx. \label{energy-3}
\end{eqnarray}
Here $\mathcal{H}_{l}$ is the kinetic energy (related to dispersion), $\mathcal{H}_{nl}$ is the potential energy (related to nonlinearity), and $\psi_{k}$ is the Fourier-transformed wave field,
$$
	\psi_{k}(t) = \mathcal{F}[\psi] = \frac{1}{L}\int_{-L/2}^{L/2}\psi(x,t)\,e^{-ikx}\,dx.
$$
When modeling the time evolution, our numerical scheme conserves the first $10$ integrals~(\ref{integrals_rec1})-(\ref{integrals_rec2}) up to the relative errors from $10^{-10}$ (the first three invariants) to $10^{-6}$ (the tenth invariant) orders. 

We examine the following statistical functions: the ensemble-averaged kinetic $\langle\mathcal{H}_{l}\rangle$ and potential $\langle\mathcal{H}_{nl}\rangle$ energies, the kurtosis $\kappa_{4}=\langle\overline{|\psi|^{4}}\rangle/\langle\overline{|\psi|^{2}}\rangle^{2}$, the probability density function (PDF) $\mathcal{P}(I)$ of relative wave intensity $I=|\psi|^{2}/\langle\overline{|\psi|^{2}}\rangle$ where $\langle\overline{|\psi|^{2}}\rangle$ is the average intensity, the Fourier spectrum,
\begin{equation}\label{wave-action-spectrum}
	S_{k} = \frac{\langle|\psi_{k}|^{2}\rangle}{\Delta k},
\end{equation}
where $\Delta k = 2\pi/L$ is distance between neighbor wavenumbers, and the autocorrelation of the intensity,
\begin{equation}\label{g2}
	g^{(2)}(x) = \frac{\langle \overline{|\psi(y+x)|^{2}\cdot |\psi(y)|^{2}}\rangle}{\langle \overline{|\psi(y)|^{2}}\rangle^{2}}.
\end{equation}
In the latter relation, the overline denotes spatial averaging over the $y$ coordinate. 
Note that, at $x=0$, the autocorrelation equals the kurtosis, $g^{(2)}(0)=\kappa_{4}$, and at $|x|\to\infty$ it must approach unity, $g^{(2)}(x)\to 1$.
For the Fourier spectrum and the PDF, we use normalization conditions $\int S_{k}\,dk = \mathcal{N}$ and $\int \mathcal{P}(I)\,dI = 1$, respectively.

The $128$-SS occupy only about two-thirds of the simulation box $x\in [-\tilde{L}/2, \tilde{L}/2]$, $\tilde{L} = 384\pi$, and are small in the remaining one-third of it, see Fig.~\ref{fig:figCW-SG}. 
To enable direct comparison with the MI case and to avoid edge effects, for the soliton gas case we additionally renormalize the Fourier spectrum to the portion of the box $r=L_{N}/\tilde{L}$, which is occupied by the $128$-SS, 
\begin{equation}\label{wave-action-spectrum-SG}
	S_{k}^{\mathrm{SG}} = \frac{\langle|\psi_{k}|^{2}\rangle}{r\,\Delta k},
\end{equation}
and compute the other statistical functions within a smaller central part of the simulation box $x\in [-\ell, \ell]$, $\ell = 250$. 
To additionally verify our results on the Fourier spectrum, we also consider the spatial correlation function,
\begin{equation}
	g^{(1)}(x) = \frac{\langle\overline{\psi(y+x)\psi^{*}(y)}\rangle}{\langle\overline{|\psi(y)|^{2}}\rangle},
	\label{correlation-function}
\end{equation}
where the overline denotes spatial averaging over the $y$ coordinate; for periodic wave fields, this function is related to the Fourier spectrum via the inverse Fourier transform, 
\begin{equation}
	g^{(1)}(x) = \frac{\Delta k}{\mathcal{N}}\,\mathcal{F}^{-1}[S_{k}], \quad \mathcal{N} = \langle\overline{|\psi|^{2}}\rangle.
	\label{correlation-function-Sk}
\end{equation}
We compute the correlation function directly according to its definition~(\ref{correlation-function}): for the soliton gas -- in the central part of the wave field $x\in [-\ell, \ell]$, while for the noise-induced MI -- over the whole simulation box. 
This allows us to compare the Fourier spectra via their proxies -- the spatial correlation functions -- without using any normalization coefficients such as $r$ in~(\ref{wave-action-spectrum-SG}). 


\section{$N$-soliton solution of fNLSE}
\label{sec:nsoliton}

The numerical algorithm employed for generating $N$-SS relies on the dressing method, see e.g.~\cite{gelash2018strongly}. 
In this formalism, multi-soliton solutions are built recursively according to the relation
\begin{equation}
    \psi_{(n)}(x) = \psi_{(n-1)}(x) + \frac{2i(\lambda_n - \lambda_n^*) q^*_{n1} q_{n2}}{| \mathbf{q}_n |^2}, \tag{15}
\end{equation}
where $\psi_{(n)}$ is the $n$-SS and $\mathbf{q}_n=\left(q_{n1},q_{n2} \right)^{\rm T}$ is the vector obtained from the scattering data $\lbrace \lambda_n, C_n\rbrace$ of the $n$-th soliton  and the solution $\mathbf{\Phi}^{(n-1)}$ of the ZS system corresponding to the $\psi_{(n-1)}$ potential,
\begin{equation}
    \mathbf{q}_n(x) = \mathbf{\Phi}^{(n-1)}(x, \lambda_n^*) \begin{pmatrix}
1 \\
C_n
\end{pmatrix}.
\end{equation}
The corresponding solution of the ZS system $\mathbf{\Phi}^{(n)}(x,\lambda)$ is computed recursively using the so-called dressing matrix $\boldsymbol{ \sigma}^{(n)}$,
\begin{equation}
    \mathbf{\Phi}^{(n)}(x, \lambda) = \boldsymbol{\sigma}^{(n)}(x, \lambda) \mathbf{\Phi}^{(n-1)}(x, \lambda),\quad 
    \sigma^{(n)}_{ml}(x, \lambda) = \delta_{ml} + \frac{\lambda_n - \lambda_n^*}{\lambda - \lambda_n} \frac{q_{nm}^* q_{nl}}{| \mathbf{q}_n |^2},
\end{equation}
where $ m, l = 1, 2 $ and  $\delta_{ml}$ is the Kronecker delta. Starting from the trivial seed solution of the fNLSE $\psi(0) = 0$ and the corresponding solution of the ZS system,
\begin{equation}
    \mathbf{\Phi}^{(0)}(x,\lambda) = \begin{pmatrix}
e^{-i\lambda x} & 0 \\
0 & e^{i\lambda x}
\end{pmatrix},
\end{equation}
one can construct a multi-soliton potential adding one soliton at each step. 
Note that the time dependence is encoded in the scattering data as $C_n(t) = C_n(0) e^{-2i\lambda_n^2 t}$.


\section{Numerical implementation of uniform soliton condensate}
\label{Sec:Implementation}

In order to implement realizations of spatially uniform SG numerically we invoke here the results of \cite{gurevich2000statistical} where $N$-SS of the KdV equation with random spatial phases $x_{0n}$ were considered in the semi-classical approximation ($N \gg 1$ and $x,t$ scaled as $N^{-1}$). 
These results have been successfully exploited in \cite{congy_dispersive_2023,congy2024riemannproblempolychromaticsoliton} to implement uniform KdV soliton condensate. 
Although similar results have not been derived for the fNLSE, they have been formally adapted in \cite{congy2024riemannproblempolychromaticsoliton} to successfully realize uniform, bound state soliton condensates of the fNLSE. This mapping between the KdV soliton condensates and their bound-state fNLSE counterparts is possible since in both cases the SG spectrum is located along an axis ($\lambda \in \mathbb{R}$ for KdV and $\lambda \in i\mathbb{R}$ for the bound state fNLSE solutions).

In both cases, one attains an (approximately) uniform SG in the spatial region $[-L/2, L/2]$ with $x$-independent DOS $f(\eta)$ by distributing uniformly the position parameters $x_{0n}$ on the interval 
\begin{equation} \label{Is}
    I_s = \left[-\frac{N}{2\kappa_s},\frac{N}{2\kappa_s} \right],
\end{equation}
and the imaginary parts of the fNLSE eigenvalues $\zeta_n$  according to a normalized density $\phi(\eta)$ on $\Gamma^+=[\eta_1, \eta_2]$, see Eq.~(\ref{phi}). 
Here $\kappa_s$ and $\phi(\eta)$ are expressions depending solely on the spectral scaling function $\sigma(\eta)$ entering the first NDR in \eqref{NDRs}. 
For fNLSE, multi-soliton solutions have additional degrees of freedom in the form of the phase parameters $\theta_{0n}$, which are independent random values uniformly distributed on the interval $[0,2\pi)$ as described in the main text.

As pointed out in Section \ref{sec:bound}, the first NDR \eqref{NDRs} for the bound state SG coincides with the counterpart NDR of KdV SG, and one can use the formulas for $\kappa_s$ and $\phi(\eta)$ derived within the KdV context as an ansatz for fNLSE bound state soliton gas. 
Below we summarize the key relations used in \cite{congy2024riemannproblempolychromaticsoliton} (the details and references can be found therein).

The density of the position parameters $x_{0n}$ on $I_s$ \eqref{Is} is given by 
\begin{equation}
  \label{eq:kappas}
  \kappa_s = \int_{\Gamma^+} \frac{\eta}{\sigma(\eta)} d \eta,
\end{equation}
where $\sigma(\eta)>0$ is the  spectral scaling function, which is expressed in terms of the DOS by the first  NDR \eqref{NDRs}  (see \cite{el_thermodynamic_2003, el_soliton_2021, bonnemain_generalized_2022}),
\begin{equation}
	\label{eq:sigma}
	\sigma(\eta) = \frac{1}{f(\eta)} \left(\eta-\int_{\Gamma^+}\ln \left| \frac{\eta+\mu}{\eta-\mu}\right| f(\mu) d\mu\right).
\end{equation}
The  eigenvalue parameters $\zeta_n$ are distributed on  $\Gamma^+=[\eta_1, \eta_2]$ with the density
\begin{equation}
  \label{eq:rho}
  \phi(\eta) = \frac{1}{\kappa_s} \frac{\eta}{\sigma(\eta)},
\end{equation}
where the pre-factor $1/\kappa_s$ ensures that $\phi(\eta)$ is normalized to unity.

Generally, the spatial density of solitons $\kappa = \int_{\Gamma^+} f(\eta) d\eta$ on $x \in [-L/2, L/2]$ does not coincide with the  density of the positional parameters $\kappa_s$ on $I_s$. For diluted soliton condensates with the DOS $f(\eta)=C f^{(1)}(\eta;\eta_1,\eta_2)$, where  $0<C<1$ one has on using the identity $\int_{\Gamma^+}\ln \big| \frac{\eta+\mu}{\eta-\mu}\big|f^{(1)}(\mu) d\mu=\eta$:
\begin{equation}
\label{eq:density_cond}
 \kappa_s= \frac{\kappa}{1-C} = \frac{C \kappa^{(1)}}{1-C},\quad \phi(\eta) = \frac{f(\eta)}{\kappa} = \frac{f^{(1)}(\eta)}{\kappa^{(1)}} ,\quad
    \kappa^{(1)} = \int_{\Gamma^+} f^{(1)}(\eta;\eta_1,\eta_2) d\eta = U(\eta_2),
\end{equation}
cf. \eqref{phi}, \eqref{eq:U2}.  For the soliton condensate ($C=1$), $\kappa_s$ diverges and $I_s$ reduced to $\{0\}$, i.e. all the position parameters are set to $0$.


\section{Useful integrals}
\label{Sec:Integrals}

Below we present several useful integrals used in the evaluation of the ensemble averages \eqref{moments24},
\begin{align}
    &\int_{\eta_1}^{\eta_2} \frac{\eta\,d\eta}{\pi \sqrt{(\eta^2-\eta_1^2)(\eta_2^2-\eta^2)}} = \frac12, \nonumber\\
    &\int_{\eta_1}^{\eta_2} \frac{\eta^3\,d\eta}{\pi \sqrt{(\eta^2-\eta_1^2)(\eta_2^2-\eta^2)}} = \frac{1}{4} \left(\eta_1^2+\eta_2^2\right), \nonumber \\
    &\int_{\eta_1}^{\eta_2} \frac{\eta^5\,d\eta}{\pi \sqrt{(\eta^2-\eta_1^2)(\eta_2^2-\eta^2)}} = \frac{1}{16} \left(3 \eta_1^4+2 \eta_2^2 \eta_1^2+3 \eta_2^4\right). \nonumber
\end{align}


\end{document}